# Why is the magnetic force similar to a Coriolis force?

Antoine Royer

*Département de Physique, Université de Montréal, Montréal, Québec H3C 3J7, Canada*[*]

It is pointed out that the underlying reason why the magnetic force is similar to a Coriolis force is that it is caused by Thomas rotations, induced by successions of non-collinear Lorentz boosts. The magnetic force may even be viewed as a kind of Coriolis force (making perhaps more acceptable the apparent non-existence of magnetic monopoles). We also show that under a change of inertial frames, Faraday lines of force Lorentz contract as if 'etched' in space, while 'Coriolis' terms get added on.

## 1 Introduction

> *What led me more or less directly to the special theory of relativity was the conviction that the electromotive force acting on a body in motion in a magnetic field was nothing else but an electric field.*   A. Einstein [1]
>
> *This magnetic force has a strange directional character* […] *Magnetism is in reality a relativistic effect of electricity.*   R. P. Feynman [2a]

As is well known, Lorentz covariance implies that a purely electric force in some inertial frame $S_0$ acquires a magnetic component in another inertial frame $S$ [2-12]. This can be seen physically in various ways. For instance, an axially moving current-carrying wire exerts a radial force on a test charge *at rest*, because the density of conduction electrons, and that of positive charges, equal in the wire's rest frame, do not stay so in the moving wire due to different Lorentz contractions of the volumes they occupy; but the radial force on the test charge is the same in both frames (to order $v^2/c^2$) [2b, 8a, 9a].

This pretty argument, however, does not work if the test charge moves transversally to the wire. Also, it does not reveal the underlying reason for the *"strange directional character"* of the magnetic force, perpendicular to the velocity, hence doing no work, similarly to a Coriolis force – of *kinematic* (rather than dynamic) origin.

We here point out that magnetic forces have their strange Coriolis-like character because they enact Thomas rotations, induced by successions of non-collinear boosts [10-14]. So if one views magnetism as a *"relativistic effect of electricity"* (Feynman), as was apparently also the initial idea of Einstein (citations above), then magnetic forces *are* a kind of Coriolis force. More generally, any *Newtonian* force (i.e. depending on the positions, but not on the velocities of particles) in some inertial frame, is necessarily accompanied by a 'Coriolis' force in another inertial frame.

---

[*] antoine.royer@umontreal.ca

Let us first recall that, as was implicitly noted by Einstein in his founding 1905 article on special relativity [3], a succession of two non-collinear Lorentz boosts does not result in a pure boost, but in a boost times a rotation, usually called a Thomas (or Wigner) rotation [13, 14].

Now, an electric field $\mathbf{e}_0$ in inertial frame $S_0$ induces, on a charged particle, an infinitesimal boost during an infinitesimal time interval. In going from $S_0$ to another inertial frame $S$, this infinitesimal boost gets Lorentz transformed. Thereby non-collinear boosts get combined, whence an infinitesimal Thomas rotation. It results that an infinitesimal boost induced by $\mathbf{e}_0$ in $S_0$ becomes in $S$ an infinitesimal boost times an infinitesimal rotation. The latter rotates the velocity of the particle in $S$, and this rotation is attributed to a 'magnetic' field $\mathbf{b}$. Yet what is acting is *"nothing else but an electric field"* [1].

The point of view that *"magnetism is in reality a relativistic effect of electricity"* [2a] is not the most economical for the mind. It is usually more efficient to think of electric and magnetic fields as equally 'real' components of a 2-tensor. On the other hand, it is often useful to look at mathematical equations from different perspectives, and to have different physical pictures and interpretations, even if this has no consequences in practice.

Observe, however, that different physical interpretations may lead to different physical *expectations*. Thus, if one views magnetism as a 'fictive' relativistic 'Coriolis' effect, then one does not expect magnetic (i.e., 'Coriolis') monopoles to exist – as is apparently the case (for now). [1] In the 'equally real' point of view, on the contrary, they would be most welcome: Classically, magnetic charges make Maxwell's equations totally symmetric [15a]; while in quantum mechanics, which *requires* magnetic potentials (incompatible, classically, with magnetic charges), the mere existence of a *single* magnetic monopole would imply (hence explain) the quantization of electric charges [16-18].

---

[1] It is interesting that Feynman [2c] asserts flatly:*"There are no magnetic "charges". None has been found"*, a bias perhaps due to his viewing magnetism as merely *"a relativistic effect of electricity"*.

Of course, the view that magnetism is a 'fictive' relativistic 'Coriolis' effect only makes magnetic monopoles *seem* less likely, but in no way forbids them. Indeed, quantum mechanics often preys on 'fictive' classical entities and gives them 'reality'. For instance, magnetic potentials, long considered 'fictive', [2] appear quite 'real' in the Bohm-Aharonov effect [2d].

*The electromagnetic field*: If the magnetic force is a relativistic 'Coriolis' effect, so should the magnetic field $\mathbf{b}$ (to which it is convenient to ascribe this force). Indeed, let $\mathbf{e}_0$ be a Coulomb field in inertial frame $S_0$, obeying the electrostatic equations ($\partial_t \equiv \partial/\partial t$)

$$\partial_{t_0}\mathbf{e}_0 = 0 \text{ (static)}, \qquad \nabla_0 \cdot \mathbf{e}_0 = \rho_0 \text{ (inverse square)}, \qquad \nabla_0 \times \mathbf{e}_0 = 0 \text{ (conservative)} \qquad (1)$$

Then Lorentz transforming from $S_0$ to $S$ yields Maxwell's equations

$$c^2 \nabla \times \mathbf{b} - \partial_t \mathbf{e} = \mathbf{j}, \qquad \nabla \cdot \mathbf{e} = \rho, \qquad \nabla \cdot \mathbf{b} = 0, \qquad \nabla \times \mathbf{e} + \partial_t \mathbf{b} = 0 \qquad (2)(a,b,c,d)$$

wherein $\mathbf{b}$ is again due to Thomas rotations. Thus Faraday's magnetic induction law (d) (a changing magnetic field creates an electric field), the other facet of magnetism alongside the magnetic force, emerges. Interestingly, Ampere's law (a), *including* Maxwell's correction $\partial_t \mathbf{e}$ (a changing electric field creates a magnetic field), is also generated by a *Galilean* transformation, but with no physical effect, for no magnetic induction, nor a magnetic force, are produced thereby.

Electric and magnetic fields, $\mathbf{e}$ and $\mathbf{b}$, boost and rotate the velocities of charged particles (making these fields detectable in the first place); that is, they Lorentz transform these velocities, hence act as *Lorentz generators*. So by analysing how an infinitesimal Lorentz transformation, *inside* some inertial frame $S$, itself transforms in going to $S'$, one can deduce, and understand *kinematically*, the way $\mathbf{e}$ and $\mathbf{b}$ Lorentz transform, namely [2-12]:

$$\mathbf{e}' = \mathbf{e}_\| + \gamma_v \mathbf{e}_\perp + \gamma_v \mathbf{b} \times \mathbf{v}, \qquad \mathbf{b}' = \mathbf{b}_\| + \gamma_v \mathbf{b}_\perp - c^{-2}\gamma_v \mathbf{e} \times \mathbf{v} \qquad (3)(a,b)$$

Here, $\mathbf{e}_\|$ and $\mathbf{e}_\perp$ are the components parallel and perpendicular, respectively, to the velocity $\mathbf{v}$ of frame $S$ relative to $S'$; $c$ is the velocity of light, and $\gamma_v = (1 - v^2/c^2)^{-1/2}$. We will see that $\gamma_v \mathbf{b} \times \mathbf{v}$ in (3)(a) is an 'anti-Coriolis' term, arising because $\mathbf{b}$ rotates velocities (it also appears under a Galilean transformation – without $\gamma_v$ due to length contraction), and is in fact responsible for the term $\partial_t \mathbf{e}$ in (2)(a). Its 'dual' $c^{-2}\gamma_v \mathbf{e} \times \mathbf{v}$ in (3)(b) is the 'Coriolis' magnetic term due to Thomas rotations (vanishing as $c \to \infty$), and gives rise to $\partial_t \mathbf{b}$ in (2)(d). As to the combinations $\mathbf{e}_\| + \gamma_v \mathbf{e}_\perp$ and $\mathbf{b}_\| + \gamma_v \mathbf{b}_\perp$, they conserve the numbers of Faraday lines of force piercing Lorentz-contracted surface elements. Thus, Eqs (3) tell us that lines of force Lorentz contract as if 'etched' in space, while '(anti)Coriolis' terms get added on. A famous special case is the contraction of the radial electric lines of force of a static charge, as it is set moving. [3] This effect, discovered by Heaviside in 1888 [15d, 19, 20], is often presented as an isolated and rather startling curiosity ("an *extraordinary* coincidence" [9b]). [4] Here we see on the contrary that everything is exactly as should be: Lorentz contraction of space *with everything in it*, plus '(anti)Coriolis' terms due to rotations.

---

[2] Heaviside called them *"metaphysical"* [15b]. For Maxwell, however, they had a direct physical meaning within his conception of the 'ether' [15c].

The propagation of electromagnetic waves has been described as "*a kind of a dance – one making the other*" [2e] between electric and magnetic fields inducing and sustaining one another. In the 'Coriolis' point of view, this *pas de deux* is between a 'real' electric field and its 'fictive' magnetic 'Coriolis' by-product. Such a symbiosis between 'real' and 'fictive' entities has precedents: For instance, in the endless precession-nutation of a frictionless spinning top, 'real' gravitational and 'fictive' Coriolis forces alternatively reanimate one another.

Let us now briefly recall the history of Thomas rotations, and discuss notation.

*Thomas rotations*: Two inertial frames $S$ and $S'$ are said to be *parallel* if their Cartesian frames are oriented such that if an observer in $S'$ sees the velocity of $S$ (relative to himself) as $\mathbf{v}$, then his friend in $S$ sees the velocity of $S'$ as $-\mathbf{v}$. Since this still allows arbitrary rotations of the Cartesian frames about $\mathbf{v}$ as axis, one must ask moreover that vectors perpendicular to $\mathbf{v}$ have identical coordinates in the two frames. [5] Let then $S$ be parallel to $S'$ and have velocity $\mathbf{v}$ in $S'$, which we denote by $S' \|_\mathbf{v} S$ (read right to left). Let in turn $S'' \|_\mathbf{u} S'$, so that $S'' \|_\mathbf{u} S' \|_\mathbf{v} S$.

---

[3] As another example, the circular magnetic lines around a current carrying electrically neutral wire, at rest, appear as ellipses to an observer moving transversally to the wire.

[4] See also [8b]. Heaviside himself found it *"rather remarkable"* that the field stays radial [15d].

[5] Even so (unless $\mathbf{v}$ is parallel to one of the Cartesian axes), the observer in $S'$ will not see, due to Lorentz contraction, the Cartesian axes of $S$ as parallel to his own axes, but rather obliquely oriented (hence not even mutually perpendicular).

If we denote by $\mathbf{u} \oplus \mathbf{v}$ the velocity of $S$ in $S''$, then it turns out that $\mathbf{v} \oplus \mathbf{u}$ is *not* equal to $\mathbf{u} \oplus \mathbf{v}$ if $\mathbf{u}$ and $\mathbf{v}$ are not collinear. In his founding 1905 article, Einstein evaluated $\mathbf{u} \oplus \mathbf{v}$, and its magnitude, noting [3]:

> "*It is worthy of remark that u and v enter into the expression for the resultant speed in a symmetrical manner*".

That is, $\mathbf{v} \oplus \mathbf{u}$ and $\mathbf{u} \oplus \mathbf{v}$ have equal magnitudes, hence differ by a rotation. This means that observers in $S$ and in $S''$ would disagree by this rotation about their relative velocity, implying that one frame is rotated relative to the other (parallelism of inertial frames is not transitive). Thus, a boost by velocity $\mathbf{v}$, followed by a boost by $\mathbf{u}$, results in a boost by $\mathbf{u} \oplus \mathbf{v}$, times a rotation. This was emphasized by Silberstein [10] in 1914. Yet these rotations are usually called Thomas rotations (especially in the infinitesimal case), or Wigner rotations (in the finite case) (after Thomas, who showed in 1926 one of their subtle effects in atomic physics [13], and Wigner, who gave in 1939 a famous analysis of the Lorentz-Poincaré group [14]).

*About notation*: To analyse the interplay of Lorentz boosts and spatial rotations, tensor notation is not practical. We will use instead vector-matrix notation in a (3-space,1-time) block form. This notation was (implicitly) used by Silberstein [10] in his treatment of Thomas rotations,[6] though once done with these rotations, he prefered quaternions [12, 21] for providing

> "*just enough 'union' to express the relativistic standpoint, and yet enough distinction not to amalgamate time and space entirely*."

We believe that the block matrix notation achieves the same end in a more intuitive way.[7] We will add to that notation spatial projectors, both to expedite calculations and to highlight the physical and geometrical meanings of quantities. This notation will allow us to make otherwise messy computations relatively simple and transparent.

---

[6] Silberstein displayed separately space and time components, using 3-vector notation, which amounts to using block matrices (Cayley's matrices were little familiar to physicists at the time).

[7] As Cayley remarked [15e]: *"I compare a quaternion formula to a pocket map – a capital thing to put in one's pocket, but which for use must be unfolded."* Quaternions gave rise to sarcastic exchanges in the closing decades of the 19th century; thus Heaviside [15f]: *"Quaternions furnish a uniquely simple and natural way of treating* quaternions. *Observe the emphasis."*

## 2 Vector-matrix notation

Bold letters will denote $3\times 3$ matrices $\mathbf{A}$, or column 3-vectors $\mathbf{a}$. The transpose $\mathbf{a}^T$ is a row 3-vector, so that $\mathbf{a}^T\mathbf{b} = \mathbf{a}\cdot\mathbf{b}$, and the 'dyad' $\mathbf{a}\mathbf{b}^T$ is a $3\times 3$ matrix. Four-dimensional space-time vectors and matrices will be written in (3-space, 1-time) block form as follows:

$$X = \begin{pmatrix} \mathbf{x} \\ t \end{pmatrix} = \begin{pmatrix} x \\ y \\ z \\ t \end{pmatrix}, \qquad M = \begin{pmatrix} \mathbf{A} & \mathbf{b} \\ \mathbf{c}^T & d \end{pmatrix} = \begin{pmatrix} A_{xx} & A_{xy} & A_{xz} & b_x \\ A_{yx} & A_{yy} & A_{yz} & b_y \\ A_{zx} & A_{zy} & A_{zz} & b_z \\ c_x & c_y & c_z & d \end{pmatrix} \qquad (4)$$

In this way, manipulations reduce to those of 2-dimensional vector-matrix algebra, except that the elements in these matrices do not commute.

*Spatial projectors*: We denote by $\boldsymbol{\pi}_\mathbf{u}$ the projector onto the 3-vector $\mathbf{u}$, and by $\bar{\boldsymbol{\pi}}_\mathbf{u}$ that onto planes perpendicular to $\mathbf{u}$, that is ($u$ is the length of vector $\mathbf{u}$):

$$\boldsymbol{\pi}_\mathbf{u} = u^{-2}\mathbf{u}\mathbf{u}^T = \boldsymbol{\pi}_\mathbf{u}^T, \qquad \bar{\boldsymbol{\pi}}_\mathbf{u} = \mathbf{1} - \boldsymbol{\pi}_\mathbf{u} = \bar{\boldsymbol{\pi}}_\mathbf{u}^T \qquad (5)$$

These satisfy $\boldsymbol{\pi}_\mathbf{u}^2 = \boldsymbol{\pi}_\mathbf{u}$, $\bar{\boldsymbol{\pi}}_\mathbf{u}^2 = \bar{\boldsymbol{\pi}}_\mathbf{u}$ and $\boldsymbol{\pi}_\mathbf{u}\bar{\boldsymbol{\pi}}_\mathbf{u} = \bar{\boldsymbol{\pi}}_\mathbf{u}\boldsymbol{\pi}_\mathbf{u} = 0$, so that, for scalars $a, b$:

$$(a\boldsymbol{\pi}_\mathbf{u} + b\bar{\boldsymbol{\pi}}_\mathbf{u})(a'\boldsymbol{\pi}_\mathbf{u} + b'\bar{\boldsymbol{\pi}}_\mathbf{u}) = aa'\boldsymbol{\pi}_\mathbf{u} + bb'\bar{\boldsymbol{\pi}}_\mathbf{u}, \qquad (a\boldsymbol{\pi}_\mathbf{u} + b\bar{\boldsymbol{\pi}}_\mathbf{u})^{-1} = (a^{-1}\boldsymbol{\pi}_\mathbf{u} + b^{-1}\bar{\boldsymbol{\pi}}_\mathbf{u}) \qquad (6)$$

Components parallel and perpendicular to $\mathbf{u}$, namely $\mathbf{a}_\| \equiv \boldsymbol{\pi}_\mathbf{u}\mathbf{a}$ and $\mathbf{a}_\perp \equiv \bar{\boldsymbol{\pi}}_\mathbf{u}\mathbf{a}$, satisfy:

$$\mathbf{a}_\| \times \mathbf{b}_\| = 0, \qquad (\mathbf{a}\times\mathbf{b})_\| = \mathbf{a}_\perp \times \mathbf{b}_\perp, \qquad (\mathbf{a}\times\mathbf{b})_\perp = \mathbf{a}_\perp \times \mathbf{b}_\| + \mathbf{a}_\| \times \mathbf{b}_\perp \qquad (7)$$

*Infinitesimal rotations*: Let $\varepsilon$ be an infinitesimal scalar. To rotate a 3-vector $\mathbf{u}$ by angle $\varepsilon b$ about axis $\mathbf{b}$, one adds to $\mathbf{u}$ a perpendicular vector $\varepsilon\mathbf{b}\times\mathbf{u}$. So a $3\times 3$ infinitesimal rotation matrix $\mathbf{R}_{\varepsilon\mathbf{b}}$ may be defined by $\mathbf{R}_{\varepsilon\mathbf{b}}\mathbf{u} = \mathbf{u} - \varepsilon\mathbf{b}\times\mathbf{u}$, for any $\mathbf{u}$, that is:

$$\mathbf{R}_{\varepsilon\mathbf{b}} = \mathbf{1} - \varepsilon\mathbf{J}_\mathbf{b}, \qquad \mathbf{J}_\mathbf{b} \equiv \mathbf{b}\times \qquad (8)$$

where the antisymmetric matrix $\mathbf{J}_\mathbf{b}$ can be read off from

$$\mathbf{J}_\mathbf{b}\mathbf{u} = \mathbf{b}\times\mathbf{u} = \begin{pmatrix} b_y u_z - b_z u_y \\ b_z u_x - b_x u_z \\ b_x u_y - b_y u_x \end{pmatrix} = \begin{pmatrix} 0 & -b_z & b_y \\ b_z & 0 & -b_x \\ -b_y & b_x & 0 \end{pmatrix}\begin{pmatrix} u_x \\ u_y \\ u_z \end{pmatrix}, \qquad \mathbf{J}_\mathbf{b} = -\mathbf{J}_\mathbf{b}^T \qquad (9)$$

The standard identities $(\mathbf{a}\times\mathbf{b})\times\mathbf{c} = \mathbf{c}\times(\mathbf{b}\times\mathbf{a}) = \mathbf{b}(\mathbf{a}\cdot\mathbf{c}) - \mathbf{a}(\mathbf{b}\cdot\mathbf{c}) = (\mathbf{b}\mathbf{a}^T - \mathbf{a}\mathbf{b}^T)\mathbf{c}$ imply that

$$\mathbf{J}_{\mathbf{a}\times\mathbf{b}} = \mathbf{b}\mathbf{a}^T - \mathbf{a}\mathbf{b}^T, \qquad \mathbf{J}_\mathbf{c}\mathbf{J}_\mathbf{b} = \mathbf{b}\mathbf{c}^T - (\mathbf{b}\cdot\mathbf{c})\mathbf{1} \qquad (10)(a,b)$$

## 3  Coriolis forces

We wish here to underline some aspects of Coriolis forces. Let a Cartesian frame $S_\omega$ rotate at a uniform angular frequency $\boldsymbol{\omega}$ relative to an inertial frame $S_0$. Then a trajectory $\mathbf{x}(t)$ in $S_\omega$ becomes $\mathbf{x}_0(t) = \mathbf{R}(t)\mathbf{x}(t)$ in $S_0$, where the rotation matrix $\mathbf{R}(t)$ satisfies $\dot{\mathbf{R}} = \mathbf{J}_\omega \mathbf{R}$ (overdots indicate time derivatives). Hence, $\dot{\mathbf{x}}_0 = \mathbf{R}\dot{\mathbf{x}} + \mathbf{J}_\omega \mathbf{R}\mathbf{x}$, and $\ddot{\mathbf{x}}_0 = (\mathbf{R}\ddot{\mathbf{x}} + \mathbf{J}_\omega \mathbf{R}\dot{\mathbf{x}}) + (\mathbf{J}_\omega \mathbf{R}\dot{\mathbf{x}} + \mathbf{J}_\omega^2 \mathbf{R}\mathbf{x})$, or letting $\mathbf{R} = \mathbf{1}$ at the instant considered [22]:

$$\dot{\mathbf{x}}_0 = \dot{\mathbf{x}} + \boldsymbol{\omega} \times \mathbf{x}, \qquad \ddot{\mathbf{x}}_0 = \ddot{\mathbf{x}} + 2\boldsymbol{\omega} \times \dot{\mathbf{x}} + \boldsymbol{\omega} \times (\boldsymbol{\omega} \times \mathbf{x}) \qquad (11)$$

The velocity dependent term $2\boldsymbol{\omega} \times \dot{\mathbf{x}}$ is the Coriolis acceleration. The factor 2 stems from two different effects: (i) rotation of the velocity $\dot{\mathbf{x}}$, and (ii) changes in the tangential velocity $\mathbf{v}_{tan} = \boldsymbol{\omega} \times \mathbf{x}$ due to changes in $\mathbf{x}$. The centripetal acceleration $\boldsymbol{\omega} \times (\boldsymbol{\omega} \times \mathbf{x})$ comes from the rotation of $\mathbf{v}_{tan}$. If now a Newtonian force $\mathbf{f}_0$ acts on a particle of mass $m$ in $S_0$, so that $m\ddot{\mathbf{x}}_0 = \mathbf{f}_0$, then its equation of motion inside $S_\omega$ is, by (11):

$$m\ddot{\mathbf{x}} = \mathbf{f}_0 + \mathbf{f}_{cent} + \mathbf{f}_{cor}, \qquad \mathbf{f}_{cent} = -m\boldsymbol{\omega} \times (\boldsymbol{\omega} \times \mathbf{x}), \qquad \mathbf{f}_{cor} = -2m\boldsymbol{\omega} \times \dot{\mathbf{x}} \qquad (12)$$

where $\mathbf{f}_{cent}$ and $\mathbf{f}_{cor}$ are 'fictive forces'. For instance, if you sit on a *frictionless* turntable $S_\omega$, then you follow a curved path relative to the turntable rotating under you, and you can view this curving relative to $S_\omega$ as caused by these 'forces'. If, however, there is friction, and you manage to crawl along a straight path $\mathbf{x}(t)$ on the turntable, then you do feel these 'fictive' forces (here better called 'inertial'). How can you detect that you are on a rotating frame? By *varying*, at the *same* position, your velocity $\dot{\mathbf{x}}$ relative to $S_\omega$, for the sideways Coriolis force will then vary. Otherwise, all forces seem 'real' (i.e., to depend only on position). [8]

You do not feel Coriolis forces just on a merry-go-round. Actually, the most common manifestation of Coriolis forces is the gyroscopic effect in spinning objects like tops and wheels. This is usually discussed in terms of angular momentum and torque, which is more efficient mathematically [22]. But as emphasized by Feynman [23], the gyroscopic effect is really due to

---

[8] You cannot tell fictive from real forces if your velocity $\dot{\mathbf{x}}$ at $\mathbf{x}$ on the turntable is always the same: so if you were told that it turns at rate $\Omega$, then you would call $-m\Omega \times (\Omega \times \mathbf{x}) - 2m\Omega \times \dot{\mathbf{x}}$ fictive, and all the rest 'real'. But by varying $\dot{\mathbf{x}}$, you would find that only $\Omega = \omega$ yields a 'real' force independent of $\dot{\mathbf{x}}$.

Coriolis forces felt by the individual particles constituting the spinning object.[9] In some cases, Coriolis and 'real' forces can sustain one another in an endless oscillation, as in the precession-nutation of a frictionless spinning top (see section 12).

## 4  Galilean relativity

To better understand how Lorentz transformations interact with electric and magnetic fields, it will help to first examine which aspects result from simple Galilean relativity, where our intuition is more at ease.

*Galilean transformations*: A general Galilean transformation $\Gamma$ may be written as a boost $\mathbf{x} \to \mathbf{x} + \mathbf{v}t = \mathbf{B_v}\mathbf{x}$ by velocity $\mathbf{v}$, times a spatial rotation $\mathbf{R}$:

$$\mathbf{x} \to \mathbf{x}' = \Gamma\mathbf{x} = \mathbf{R}(\mathbf{x} + \mathbf{v}t), \qquad \Gamma = \mathbf{R}\mathbf{B_v} \qquad (13)$$

Let $\varepsilon$ be an infinitesimal, and consider, using (8), an infinitesimal Galilean transformation $\mathbf{x} \to \hat{\mathbf{x}} = \mathbf{R}_{\varepsilon\mathbf{b}}\mathbf{B}_{\varepsilon\mathbf{e}}\mathbf{x} = (1 - \varepsilon\mathbf{J_b})(\mathbf{x} + \varepsilon\mathbf{e}t)$ *inside* inertial frame $S$ (we use $\mathbf{e}, \mathbf{b}$ because electric and magnetic fields act in this way on the *velocities* of charged particles). To first order in $\varepsilon$:

$$\hat{\mathbf{x}} = \mathbf{R}_{\varepsilon\mathbf{b}}\mathbf{B}_{\varepsilon\mathbf{e}}\mathbf{x} = \mathbf{x} - \varepsilon\mathbf{J_b}\mathbf{x} + \varepsilon\mathbf{e}t = \mathbf{R}_{\varepsilon\mathbf{b}}\mathbf{x} + \varepsilon\mathbf{e}t \qquad (14)(a,b,c)$$

Let now $S$ have velocity $\mathbf{v}$ relative to $S'$ and be parallel to it, so that $\mathbf{x}' = \mathbf{x} + \mathbf{v}t$. Then in $S'$: $\hat{\mathbf{x}}' = \mathbf{R}_{\varepsilon\mathbf{b}'}\mathbf{B}_{\varepsilon\mathbf{e}'}\mathbf{x}' = \mathbf{x}' - \varepsilon\mathbf{J_{b'}}\mathbf{x}' + \varepsilon\mathbf{e}'t = \mathbf{R}_{\varepsilon\mathbf{b}'}\mathbf{x}' + \varepsilon\mathbf{e}'t$ (*)(a,b,c) with *other* vectors $\mathbf{e}', \mathbf{b}'$. Inserting $\mathbf{x}' = \mathbf{x} + \mathbf{v}t$ in (*)(b), and (14)(b) in $\hat{\mathbf{x}}' = \hat{\mathbf{x}} + \mathbf{v}t$, and equating, yields $\varepsilon\mathbf{e}'t - \varepsilon\mathbf{J_{b'}}\mathbf{x} - \varepsilon\mathbf{J_{b'}}\mathbf{v}t = \varepsilon\mathbf{e}t - \varepsilon\mathbf{J_b}\mathbf{x}$. Since $\mathbf{x}$ is arbitrary, we must have $\mathbf{b}' = \mathbf{b}$. Whence:

$$\mathbf{b}' = \mathbf{b}, \qquad \mathbf{e}' = \mathbf{e} + \mathbf{J}_b\mathbf{v} = \mathbf{e} + \mathbf{b} \times \mathbf{v} \qquad (15)$$

where $\mathbf{J}_b\mathbf{v}$ in $\mathbf{e}'$ is needed to cancel out $-\varepsilon\mathbf{J_b}\mathbf{v}t$ in $\mathbf{R}_{\varepsilon\mathbf{b}}\mathbf{x}' = \mathbf{R}_{\varepsilon\mathbf{b}}(\mathbf{x} + \mathbf{v}t)$ [in (*)(c)], since $\mathbf{R}_{\varepsilon\mathbf{b}}$ in (14)(c) acts on $\mathbf{x}$ alone, and *not* on $\mathbf{v}$. We shall call $\mathbf{J}_b\mathbf{v}$ an 'anti-Coriolis' term.[10]

---

[9] Consider a spinning vertical bicycle wheel: Particles at its front end have large downwards velocities, so that *inclining* the wheel to the right forces these particles to acquire leftwards velocity components, whence Coriolis forces to the right. Particles at the back end of the wheel have upwards velocities, hence exert Coriolis forces to the left. Whence a torque turning the wheel to the right.

[10] Put otherwise, a rotation $\mathbf{R}_{\varepsilon\mathbf{b}}$ about the origin of $S$ (e.g., the center of a flying frisbee), moving with uniform velocity $\mathbf{v}$ relative to $S'$ (the ground), is equivalent to the same rotation about the origin of $S'$ (yourself say), but with an added translation $\varepsilon\mathbf{b} \times \mathbf{v}t$ *undoing* the rotation of $\mathbf{v}t$.

*Equation of motion*: Let a particle of mass $m$ have position $\mathbf{x}(t)$ and velocity $\mathbf{u} = \dot{\mathbf{x}} = d\mathbf{x}/dt$ in inertial frame $S$. Let also $\delta t$ be an infinitesimal time interval. Writing $\mathbf{u}(t+\delta t) = \mathbf{u}(t) + \dot{\mathbf{u}}\delta t$ to first order in $\delta t$, we will use Newton's law of motion $m\dot{\mathbf{u}} = \mathbf{f}$ in the form

$$\mathbf{u}(t+\delta t) = \mathbf{u}(t) + (\mathbf{f}/m)\delta t \tag{16}$$

Now, although magnetism is a Lorentzian correction of order $c^{-2}$, we can still imagine, for the sake of discussion, a magnetic force acting in a Galilean world ($c \to \infty$). So let $\mathbf{f} = q(\mathbf{e} - \mathbf{b} \times \mathbf{u})$ in (16), where $q$ is the charge on the particle, and $\mathbf{e}, \mathbf{b}$ are electric and magnetic fields in $S$. Using (8), we get, to first order in $\varepsilon \equiv q\delta t/m$ [compare (14)]:

$$\mathbf{u}(t+\delta t) = \mathbf{u}(t) + \varepsilon[\mathbf{e} - \mathbf{b} \times \mathbf{u}(t)] = \mathbf{R}_{\varepsilon\mathbf{b}}\mathbf{B}_{\varepsilon\mathbf{e}}\mathbf{u}(t), \qquad \varepsilon \equiv q\delta t/m \tag{17}$$

namely an infinitesimal Galilean transformation on $\mathbf{u}(t)$: During $\delta t$, the electric field $\mathbf{e}$ boosts the velocity $\mathbf{u}$ by $\varepsilon\mathbf{e}$, and the magnetic field $\mathbf{b}$ rotates it by $\mathbf{R}_{\varepsilon\mathbf{b}}$. Thus, $\mathbf{e}, \mathbf{b}$ act here as *Galilean generators*. So the way they transform, namely Eqs (15), is just kinematics. Note that by (15), and since the particle's velocity in $S'$ is $\mathbf{u}' = \mathbf{u} + \mathbf{v}$, we have $\mathbf{e}' - \mathbf{b}' \times \mathbf{u}' = \mathbf{e} - \mathbf{b} \times \mathbf{u}$, i.e., the particle feels the same force in both frames, $\mathbf{f}' = \mathbf{f}$, as is evident.

Observe that if $\mathbf{e} = 0$ in $S$, but $\mathbf{b} \neq 0$, then one gets $\mathbf{e}' = \mathbf{J}_\mathbf{b}\mathbf{v}$ in $S'$, that is: magnetic forces in a Galilean world necessarily imply 'anti-Coriolis' electric forces. However, electric forces do not imply magnetic forces (since $\mathbf{b}' = \mathbf{b}$), so that electric forces, that is, *Newtonian* forces, can exist solitarily in a Galilean world (unlike in our Lorentzian world).

*Field equations*: We next examine how the electrostatic field equations transform under a Galilean transformation. Coulomb's inverse square force law, and the superposition principle, imply that a *static* charge density $\rho_0$, in inertial frame $S_0$, creates an electric field $\mathbf{e}_0$ satisfying the electrostatic equations (1) (Coulomb's law in differential form):

$$\partial_{t_0}\mathbf{e}_0 = 0, \qquad \nabla_0 \cdot \mathbf{e}_0 = \rho_0, \qquad \nabla_0 \times \mathbf{e}_0 = 0 \tag{18}$$

It is perfectly legitimate to submit these equations (which also describe Newtonian gravity) to a Galilean transformation. So let $S_0$ have velocity $\mathbf{v}_0$ in $S$, so that $\mathbf{x} = \mathbf{x}_0 + \mathbf{v}_0 t$. Then $\mathbf{e}(\mathbf{x},t) = \mathbf{e}_0(\mathbf{x} - \mathbf{v}_0 t)$, and we get in $S$:

$$\partial_t \mathbf{e} + (\mathbf{v}_0 \cdot \nabla)\mathbf{e} = 0, \qquad \nabla \cdot \mathbf{e} = \rho, \qquad \nabla \times \mathbf{e} = 0 \tag{19}$$

where $\mathbf{e} = \mathbf{e}_0$ and $\rho = \rho_0$. Noting next [11] that $(\mathbf{v}_0 \cdot \nabla)\mathbf{e}_0 = \mathbf{v}_0(\nabla \cdot \mathbf{e}_0) - \nabla \times (\mathbf{v}_0 \times \mathbf{e}_0)$ and $\mathbf{v}_0(\nabla \cdot \mathbf{e}_0) = \mathbf{v}_0 \rho_0 = \mathbf{j}$, we obtain (the reason for the tilde on $\tilde{\mathbf{b}}$ will appear in section 9):

$$\nabla \times \tilde{\mathbf{b}} - \partial_t \mathbf{e} = \mathbf{j}, \qquad \nabla \cdot \mathbf{e} = \rho, \qquad \nabla \cdot \tilde{\mathbf{b}} = 0, \qquad \nabla \times \mathbf{e} = 0 \qquad (20)(a,b,c,d)$$

$$\mathbf{e} = \mathbf{e}_0, \qquad \tilde{\mathbf{b}} \equiv \mathbf{v}_0 \times \mathbf{e}_0, \qquad \rho = \rho_0, \qquad \mathbf{j} = \mathbf{v}_0 \rho_0 \qquad (21)$$

[we noted that $\nabla \cdot \tilde{\mathbf{b}} = -\mathbf{v}_0 \cdot (\nabla_0 \times \mathbf{e}_0) = 0$]. Eqs (20) have the advantage over (19) that they do not refer explicitly to $\mathbf{v}_0$. So if there are charges of various velocities $\mathbf{v}_i$ in $S$, then one can invoke the superposition principle and put (if $\mathbf{e}_i$ is due to charge $\rho_i$ in its rest frame $S_i$):

$$\mathbf{j} = \sum_i \mathbf{v}_i \rho_i, \qquad \mathbf{e}(\mathbf{x},t) = \sum_i \mathbf{e}_i(\mathbf{x}_i - \mathbf{v}_i t), \qquad \tilde{\mathbf{b}} = \sum_i \mathbf{v}_i \times \mathbf{e}_i \qquad (22)$$

The fields $\mathbf{e}, \tilde{\mathbf{b}}$ can now vary arbitrarily in both space and time. If frame $S$ has velocity $\mathbf{v}$ in $S'$, then $\mathbf{e}'(\mathbf{x}',t) = \mathbf{e}(\mathbf{x}' - \mathbf{v}t, t)$, so that $\partial_t \mathbf{e}' = \dot{\mathbf{e}} - (\mathbf{v} \cdot \nabla)\mathbf{e}$. Using $\dot{\mathbf{e}} = \nabla \times \tilde{\mathbf{b}} - \mathbf{j}$, and again $(\mathbf{v} \cdot \nabla)\mathbf{e} = \mathbf{v}(\nabla \cdot \mathbf{e}) - \nabla \times (\mathbf{v} \times \mathbf{e})$ (and $\mathbf{e} = \mathbf{e}_0$), we get:

$$\partial_t \mathbf{e}' = \nabla \times \tilde{\mathbf{b}}' - \mathbf{j}', \qquad \mathbf{j}' = (\mathbf{v}_0 + \mathbf{v})\rho_0, \qquad \mathbf{e}' = \mathbf{e}, \qquad \tilde{\mathbf{b}}' = \tilde{\mathbf{b}} + \mathbf{v} \times \mathbf{e} \qquad (23)$$

Observe that (20)(a) is Ampère's law, *including* Maxwell's correction $\partial_t \mathbf{e}$: a changing $\mathbf{e}$ induces $\tilde{\mathbf{b}}$. However, since $\nabla \times \mathbf{e} = 0$, there is no magnetic induction, so that $\tilde{\mathbf{b}}$ has no physical effect. [12] Also, since the transformation of $\mathbf{e}, \mathbf{b}$ in (23) differs from that in (15) (so that $\tilde{\mathbf{b}}$ is *different* from $\mathbf{b}$), we see that a Lorentz force $\mathbf{f} = q(\mathbf{e} - \mathbf{b} \times \mathbf{u})$, and Coulomb electrostatics (18) (or Newtonian gravity), are *separately* compatible with Galilean relativity, but *not* together (unless $\mathbf{b} = 0$). It was of course the incompatibility of Maxwell's equations (and of magnetic induction) with Galilean relativity which motivated special relativity.

---

[11] We use here the identities $\nabla \times (\mathbf{a} \times \mathbf{b}) = \mathbf{a}(\nabla \cdot \mathbf{b}) - (\mathbf{a} \cdot \nabla)\mathbf{b} - \mathbf{b}(\nabla \cdot \mathbf{a}) + (\mathbf{b} \cdot \nabla)\mathbf{a}$ and $\nabla \cdot (\mathbf{a} \times \mathbf{b}) = \mathbf{b} \cdot (\nabla \times \mathbf{a}) - \mathbf{a} \cdot (\nabla \times \mathbf{b})$, see e.g. Ref [9c].

[12] Note however that $\tilde{\mathbf{b}} \equiv \mathbf{v}_0 \times \mathbf{e}$, and $\partial_t \mathbf{e} = -(\mathbf{v}_0 \cdot \nabla)\mathbf{e}$ in (19), imply $\partial_t \tilde{\mathbf{b}} = -(\mathbf{v}_0 \cdot \nabla)\tilde{\mathbf{b}} = \nabla \times (\mathbf{v}_0 \times \tilde{\mathbf{b}})$ $-\mathbf{v}_0(\nabla \cdot \tilde{\mathbf{b}}) = -\nabla \times \overline{\mathbf{e}}$ (since $\nabla \cdot \tilde{\mathbf{b}} = 0$), hence $\partial_t \tilde{\mathbf{b}} + \nabla \times \overline{\mathbf{e}} = 0$ (*), where $\overline{\mathbf{e}} = \tilde{\mathbf{b}} \times \mathbf{v}_0 = v_0^2 \overline{\pi}_{\mathbf{v}_0} \mathbf{e}_0$ since $\overline{\mathbf{e}} = (\mathbf{v}_0 \times \mathbf{e}_0) \times \mathbf{v}_0 = -\mathbf{J}_{\mathbf{v}_0}^2 \mathbf{e}_0 = v_0^2 \overline{\pi}_{\mathbf{v}_0} \mathbf{e}_0$ by (10)(b). Eq (*) is Faraday's induction law, but with $\overline{\mathbf{e}}$ *different* from $\mathbf{e}$. Thus, all the ingredients of Maxwell's equations in fact show up in the Galilean case, but not in a coherent way ($\overline{\mathbf{e}}$ and $\tilde{\mathbf{b}}$ have no physical significance here).

## 5 Special relativity, Lorentz transformations

Lorentz transformations $X \to X' = \Lambda X$ preserve space-time intervals [13]

$$c^{-2}\mathbf{x}^2 - t^2 = X^{\mathsf{T}}gX, \qquad X = \begin{pmatrix} \mathbf{x} \\ t \end{pmatrix}, \qquad g \equiv \begin{pmatrix} c^{-2}\mathbf{1} & 0 \\ 0 & -1 \end{pmatrix} \qquad (24)$$

where $g$ is the metric. Thus, $X'^{\mathsf{T}}gX' = X^{\mathsf{T}}gX$, whence the requirement:

$$\Lambda^{\mathsf{T}}g\Lambda = g \quad \Leftrightarrow \quad \Lambda^{-1} = g^{-1}\Lambda^{\mathsf{T}}g \quad \Leftrightarrow \quad \Lambda^{-1\mathsf{T}} = g\Lambda g^{-1} \qquad (25)(a,b,c)$$

Vectors transforming like $X$ are called *contravariant*. Vectors $\tilde{Y}$ such that scalars $\tilde{Y}^{\mathsf{T}}X$ are invariant are called *covariant*, and transform as

$$\tilde{Y} \to \tilde{Y}' = \Lambda^{-1\mathsf{T}}\tilde{Y}, \qquad \tilde{Y}'^{\mathsf{T}}X' = \tilde{Y}^{\mathsf{T}}X \qquad (26)$$

(since $\tilde{Y}'^{\mathsf{T}}X' = \tilde{Y}^{\mathsf{T}}\Lambda^{-1}\Lambda X$). For instance $\tilde{\partial} \equiv (\partial_x, \partial_y, \partial_z, \partial_t) = (\nabla, \partial_t)$ (where $\partial_x = \partial/\partial x$, etc.) is covariant since $\tilde{\partial}^{\mathsf{T}}X = 4$ is invariant. Contravariant 2-tensors $\mathcal{T}$, such that scalars $\tilde{X}^{\mathsf{T}}\mathcal{T}\tilde{Y}$ are invariant, transform as $\mathcal{T} \to \mathcal{T}' = \Lambda\mathcal{T}\Lambda^{\mathsf{T}}$. Thus:

$$\tilde{\partial} \equiv \begin{pmatrix} \nabla \\ \partial_t \end{pmatrix} \to \tilde{\partial}' = \Lambda^{-1\mathsf{T}}\tilde{\partial}, \qquad \mathcal{T}' = \Lambda\mathcal{T}\Lambda^{\mathsf{T}} \qquad (27)(a,b)$$

In view of (25)(c), with every contravariant vector $Y$ is associated a covariant vector $\tilde{Y} = gY$ [since $\tilde{Y}' = g\Lambda Y = (g\Lambda g^{-1})gY = \Lambda^{-1\mathsf{T}}\tilde{Y}$], and vice versa. For instance, the contravariant vector associated with $\tilde{\partial}$ in (27) is

$$\partial \equiv g^{-1}\tilde{\partial} = \begin{pmatrix} c^2\nabla \\ -\partial_t \end{pmatrix} \to \partial' = \Lambda\partial, \qquad \tilde{Y} \equiv gY \qquad (28)$$

Let inertial frames $S$ and $S'$ be related by $X' = \Lambda X$. Then a transformation $T$ inside $S$ becomes $T' = \Lambda T\Lambda^{-1}$ inside $S'$, since $Y = TX \to Y' = \Lambda Y = \Lambda TX = (\Lambda T\Lambda^{-1})\Lambda X = T'X'$. Thus, in contrast to (27)(b):

$$Y = TX \quad \to \quad Y' = T'X' \quad \text{where} \quad T' = \Lambda T\Lambda^{-1} \qquad (29)$$

Note, however, in view (25)(b), that $\mathcal{T} \equiv Tg^{-1} \to \mathcal{T}' = \Lambda\mathcal{T}\Lambda^{\mathsf{T}}$ is a contravariant tensor.

---

[13] See Appendix A for the connection with tensor notation. We use $X = (\mathbf{x}, t)$, rather than the usual $(\mathbf{x}, ct)$, in order to easily deduce the Galilean limit $c \to \infty$.

*Explicit form of Lorentz transformations*: We will denote, using the projectors (5):

$$\gamma_v \equiv (1-v^2/c^2)^{-1/2}, \qquad \mathbf{K_v} \equiv \boldsymbol{\pi_v} + \gamma_v^{-1}\bar{\boldsymbol{\pi}}_v, \qquad \mathbf{K_v} = \mathbf{K_v^T} = \mathbf{K_{-v}} \qquad (30)(a,b,c)$$

Since $\mathbf{K_v x} = \mathbf{x}_\| + \gamma_v^{-1}\mathbf{x}_\perp$, we see that $\mathbf{K_v}$ compresses transverse components by $\gamma_v^{-1} \leq 1$, hence shortens and *rotates* 3-vectors. Now, any matrix $\Lambda$ satisfying (25) can uniquely be written as a product $\Lambda = R\mathcal{B}_v$ of a spatial rotation $R$ and of a boost $\mathcal{B}_v$ [7, 10]:

$$\Lambda = R\mathcal{B}_v, \qquad \mathcal{B}_v \equiv \gamma_v \begin{pmatrix} \mathbf{K_v} & \mathbf{v} \\ \mathbf{v}^T/c^2 & 1 \end{pmatrix}, \qquad R \equiv \begin{pmatrix} \mathbf{R} & 0 \\ 0 & 1 \end{pmatrix} \qquad (31)$$

where $\mathcal{B}_v^{-1} = \mathcal{B}_{-v}$, $\mathbf{R}^{-1} = \mathbf{R}^T$. A pure (active) boost $X \to X' = \mathcal{B}_v X$ reads

$$\begin{pmatrix} \mathbf{x'} \\ t' \end{pmatrix} = \gamma_v \begin{pmatrix} \mathbf{K_v} & \mathbf{v} \\ \mathbf{v}^T/c^2 & 1 \end{pmatrix}\begin{pmatrix} \mathbf{x} \\ t \end{pmatrix} = \gamma_v \begin{pmatrix} \mathbf{K_v x} + \mathbf{v}t \\ c^{-2}\mathbf{v}^T\mathbf{x} + t \end{pmatrix} \qquad (32)$$

whence the familiar relations

$$\mathbf{x'} = \mathbf{x}_\perp + \gamma_v(\mathbf{x}_\| + \mathbf{v}t), \qquad t' = \gamma_v(t + \mathbf{v}\cdot\mathbf{x}/c^2) \qquad (33)$$

*Time-dilation and length contraction*: Applying (32) to intervals $\Delta X = (\Delta \mathbf{x}, \Delta t)$ gives us

$$\Delta \mathbf{x'} = \gamma_v(\mathbf{K_v}\Delta\mathbf{x} + \mathbf{v}\Delta t), \qquad \Delta t' = \gamma_v(\Delta t + \mathbf{v}\cdot\Delta\mathbf{x}/c^2) \qquad (34)(a,b)$$

Let us recall then the two most famous relativistic effects: Two events separated by a time interval $\Delta t$ at the same location in $S$, so that $\Delta \mathbf{x} = 0$ in (34), are separated in $S'$ by

$$\Delta t' = \gamma_v \Delta t \quad \text{if} \quad \Delta \mathbf{x} = 0 \qquad \text{(time dilation)} \qquad (35)$$

By the inverse $X = \mathcal{B}_{-v}X'$ of (32), two events distant by $\Delta \mathbf{x'}$ but simultaneous in $S'$, so that $\Delta t' = 0$, are distant by $\Delta \mathbf{x} = \gamma_v \mathbf{K_v} \Delta \mathbf{x'}$ in $S$, hence, by (6):

$$\Delta \mathbf{x'} = \gamma_v^{-1}\mathbf{K_v}^{-1}\Delta \mathbf{x} = \gamma_v^{-1}\Delta\mathbf{x}_\| + \Delta\mathbf{x}_\perp \quad \text{if} \quad \Delta t' = 0 \qquad (36)$$

exhibiting the Lorentz-Fitzgerald contraction of longitudinal lengths. Using $d\mathbf{x'} = \gamma_v^{-1}d\mathbf{x}_\| + d\mathbf{x}_\perp$ (if $dt' = 0$) and (7), we find that *surface elements* $d\mathbf{s} = d\mathbf{x} \times d\mathbf{y}$ transform as

$$d\mathbf{s'} = d\mathbf{x'} \times d\mathbf{y'} = d\mathbf{s}_\| + \gamma_v^{-1}d\mathbf{s}_\perp = \mathbf{K_v}d\mathbf{s} \qquad (37)$$

(the longitudinal dimension of the surface, i.e., the component $d\mathbf{s}_\perp$ of its normal, contracts).

*The 4-velocity*: Consider a particle of position $\mathbf{x}(t)$ and velocity $\mathbf{u}(t) = d\mathbf{x}/dt$ in inertial frame $S$: since it stays fixed ($\Delta\mathbf{x}_0 = 0$) in its rest frame $S_0$, its 'proper time' $\tau \equiv t_0 = t/\gamma_u$ [by (35)] is an invariant. It follows that the 4-velocity $U \equiv dX/d\tau$ transforms like $X$. Thus:

$$U \equiv \frac{dX}{d\tau} = \frac{d}{d\tau}\begin{pmatrix}\mathbf{x}\\t\end{pmatrix} = \gamma_u \begin{pmatrix}\mathbf{u}\\1\end{pmatrix}, \qquad U' = \Lambda U, \qquad \tau = t/\gamma_u \qquad (38)$$

## 6 Infinitesimal Lorentz transformations

Let $\varepsilon$ be an infinitesimal. Since $\gamma_{\varepsilon\mathbf{e}} = 1$ to first order in $\varepsilon$, and $\mathbf{R}_{\varepsilon\mathbf{b}} = \mathbf{1} - \varepsilon\mathbf{J}_\mathbf{b}$ by (8), an infinitesimal Lorentz transformation has the general form

$$\Lambda_{\varepsilon\mathbf{e},\varepsilon\mathbf{b}} = R_{\varepsilon\mathbf{b}}\mathcal{B}_{\varepsilon\mathbf{e}} = \mathcal{B}_{\varepsilon\mathbf{e}}R_{\varepsilon\mathbf{b}} = 1 + \varepsilon\mathcal{G}_{\mathbf{e},\mathbf{b}}, \qquad \mathcal{G}_{\mathbf{e},\mathbf{b}} \equiv \begin{pmatrix}-\mathbf{J}_\mathbf{b} & \mathbf{e}\\ \mathbf{e}^\mathsf{T}/c^2 & 0\end{pmatrix} \qquad (39)$$

where $\mathcal{G}_{\mathbf{e},\mathbf{b}}$ is a *Lorentz generator*. If inertial frames $S$ and $S'$ are related by $X' = \Lambda X$, then by (29) a transformation $1 + \varepsilon\mathcal{G}_{\mathbf{e},\mathbf{b}}$ inside $S$ becomes $1 + \varepsilon\Lambda\mathcal{G}_{\mathbf{e},\mathbf{b}}\Lambda^{-1}$ inside $S'$. Since this is also a Lorentz transformation, we must have

$$\Lambda\mathcal{G}_{\mathbf{e},\mathbf{b}}\Lambda^{-1} = \mathcal{G}_{\mathbf{e}',\mathbf{b}'} \qquad (40)$$

with transformed vectors $\mathbf{e}',\mathbf{b}'$. If $\Lambda = \mathcal{B}_\mathbf{v}$ is a pure boost, one gets (see Appendix B.1):

$$\begin{aligned}&\mathcal{B}_\mathbf{v}\mathcal{G}_{\mathbf{e},\mathbf{b}}\mathcal{B}_\mathbf{v}^{-1} = \mathcal{G}_{\mathbf{e}',\mathbf{b}'}\\ &\mathbf{e}' = \mathbf{K}_\mathbf{v}^{-1}\mathbf{e} + \gamma_v \mathbf{b}\times\mathbf{v}, \qquad \mathbf{b}' = \mathbf{K}_\mathbf{v}^{-1}\mathbf{b} - c^{-2}\gamma_v \mathbf{e}\times\mathbf{v}\end{aligned} \qquad (41)$$

[identical to (3)]. Note that the term $\gamma_v \mathbf{b}\times\mathbf{v}$ in $\mathbf{e}'$ survives in the Galilean limit $c \to \infty$ (since $\gamma_v \to 1$), and is just the 'anti-Coriolis' term $\mathbf{J}_\mathbf{b}\mathbf{v}$ in (15) (the role of $\mathbf{K}_\mathbf{v}^{-1}$ will be seen later on). We note also that transforming a *pure boost* $\mathcal{B}_{\varepsilon\mathbf{e}}$ with $\mathcal{B}_\mathbf{v}$ yields

$$\mathcal{B}_\mathbf{v}\mathcal{B}_{\varepsilon\mathbf{e}}\mathcal{B}_\mathbf{v}^{-1} = R_{\varepsilon\mathbf{b}'}\mathcal{B}_{\varepsilon\mathbf{e}'}, \qquad \mathbf{e}' = \mathbf{K}_\mathbf{v}^{-1}\mathbf{e}, \qquad \mathbf{b}' = -c^{-2}\gamma_v \mathbf{e}\times\mathbf{v} \qquad (42)$$

Here, $\mathbf{R}_{\varepsilon\mathbf{b}'} = \mathbf{1} + \varepsilon c^{-2}\gamma_v \mathbf{J}_{\mathbf{e}\times\mathbf{v}}$ is an infinitesimal *Thomas rotation*.

*Invariants and duality*: Since $\mathrm{tr}M = \mathrm{tr}\Lambda M\Lambda^{-1}$ and $\det M = \det \Lambda M\Lambda^{-1}$, the trace and determinant of powers of $\mathcal{G}$ are invariants. In particular, $\mathrm{tr}\mathcal{G}_{\mathbf{e},\mathbf{b}}^2$ and $\det\mathcal{G}_{\mathbf{e},\mathbf{b}}$ yield the following invariants (see Appendix B.2) [7]:

$$\mathbf{e}^2 - c^2\mathbf{b}^2 = \mathbf{e}'^2 - c^2\mathbf{b}'^2, \qquad \mathbf{e}\cdot\mathbf{b} = \mathbf{e}'\cdot\mathbf{b}' \tag{43}$$

If one defines a 'dual' $\bar{\mathcal{G}}_{\mathbf{e},\mathbf{b}} \equiv \mathcal{G}_{\mathbf{b},-\mathbf{e}/c^2}$, then one finds that [7]:

$$\bar{\mathcal{G}}_{\mathbf{e},\mathbf{b}} \equiv \mathcal{G}_{\mathbf{b},-\mathbf{e}/c^2} \quad\Rightarrow\quad \mathcal{G}\bar{\mathcal{G}} = (\mathbf{e}\cdot\mathbf{b}/c^2)\mathbf{1} \quad\Rightarrow\quad \bar{\mathcal{G}}_{\mathbf{e}',\mathbf{b}'} = \Lambda\bar{\mathcal{G}}_{\mathbf{e},\mathbf{b}}\Lambda^{-1} \tag{44}$$

## 7  Relativistic equation of motion

All our special relativistic considerations up to now were purely kinematic. We now turn to dynamics. We will start with a particle *at rest* in inertial frame $S_0$, and let a Newtonian force (depending only on position) act on it. Since the particle has zero velocity, Newton's law applies at time $t_0 = 0$. Lorentz transforming to another inertial frame will then yield the relativistic equation of motion. We will find that due to the Thomas rotation in (42), a velocity-rotating force arises, so that Newtonian forces *imply* Coriolis-like forces. The reciprocal is also true, as in the Galilean case (15). Hence, in Lorentzian relativity, any (rest-mass preserving) force necessarily has both an 'electric' and a 'magnetic' component [7]. This being so, we may as well assume that the Newtonian force acting in $S_0$ is an electric force.

So consider a particle of mass $m$ and electric charge $q$, *at rest* in $S_0$ at time $t_0 = \tau$ (its proper time). Its velocity is $\mathbf{u}_0(\tau) = d\mathbf{x}_0/d\tau = 0$, and its 4-velocity (38) is $U_0(\tau) = \begin{pmatrix}\mathbf{0}\\1\end{pmatrix}$. Let a Newtonian force $\mathbf{f}_0 = q\mathbf{e}_0$ act on the particle, where $\mathbf{e}_0$ is an electric field in $S_0$. Newton's law (which applies at zero velocity) implies, by (16), that an infinitesimal time $\delta\tau$ later, $\mathbf{u}_0(\tau+\delta\tau) = (\mathbf{f}_0/m)\delta\tau = \varepsilon\mathbf{e}_0$, where $\varepsilon \equiv q\delta\tau/m$. Hence, to first order in $\varepsilon$:

$$U_0(\tau+\delta\tau) = \begin{pmatrix}\varepsilon\mathbf{e}_0\\1\end{pmatrix} = \begin{pmatrix}\mathbf{1} & \varepsilon\mathbf{e}_0\\ \varepsilon\mathbf{e}_0^\mathsf{T}/c^2 & 1\end{pmatrix}\begin{pmatrix}\mathbf{0}\\1\end{pmatrix} = \mathcal{B}_{\varepsilon\mathbf{e}_0}U_0(\tau), \qquad \varepsilon \equiv q\delta\tau/m \tag{45}$$

(where $\varepsilon\mathbf{e}_0^\mathsf{T}/c^2$ could be added in since it multiplies 0). Thus, during $\delta\tau$, the 4-velocity gets boosted by $\mathcal{B}_{\varepsilon\mathbf{e}_0}$. Now if the particle (hence $S_0$) had, at its proper time $\tau$, velocity $\mathbf{u}(\tau)$ in inertial frame $S$, then its 4-velocity in $S$ at time $\tau+\delta\tau$ will be, by (42):

$$U(\tau+\delta\tau) = \mathcal{B}_\mathbf{u}U_0(\tau+\delta\tau) = (\mathcal{B}_\mathbf{u}\mathcal{B}_{\varepsilon\mathbf{e}_0}\mathcal{B}_\mathbf{u}^{-1})\mathcal{B}_\mathbf{u}U_0(\tau) = R_{\varepsilon\mathbf{b}}\mathcal{B}_{\varepsilon\mathbf{e}}U(\tau)$$
$$\mathbf{e} = \mathbf{K}_\mathbf{u}^{-1}\mathbf{e}_0, \qquad \mathbf{b} = -c^{-2}\gamma_u\,\mathbf{e}_0\times\mathbf{u} \tag{46}$$

Using (39) and $U(\tau) = \mathcal{B}_\mathbf{u}U_0(\tau) = \gamma_u\begin{pmatrix}\mathbf{u}\\1\end{pmatrix}$, we can rewrite (46) as

$$\frac{d}{d\tau}U(\tau) = \frac{d}{d\tau}\begin{pmatrix} \gamma_u \mathbf{u} \\ \gamma_u \end{pmatrix} = \frac{q\gamma_u}{m}\begin{pmatrix} \mathbf{e} - \mathbf{b} \times \mathbf{u} \\ \mathbf{e} \cdot \mathbf{u}/c^2 \end{pmatrix} \tag{47}$$

whence, since $\delta t = \gamma_u \delta \tau$, the familiar relativistic equations of motion [4-12]:

$$d\mathbf{p}/dt = q(\mathbf{e} - \mathbf{b} \times \mathbf{u}) = \mathbf{f}, \qquad dE/dt = q\mathbf{e} \cdot \mathbf{u} = \mathbf{f} \cdot \mathbf{u} \tag{48(a,b)}$$

Here, $\mathbf{p} = m\gamma_u \mathbf{u}$ and $E = m\gamma_u c^2$ are the relativistic momentum and energy, and (48)(b) reminds us that the magnetic force $\mathbf{b} \times \mathbf{u}$ does no work.

Thus, Lorentz transforming the infinitesimal boost $\mathcal{B}_{\varepsilon \mathbf{e}_0}$ inside $S_0$ to $\mathcal{B}_\mathbf{u} \mathcal{B}_{\varepsilon \mathbf{e}_0} \mathcal{B}_\mathbf{u}^{-1} = R_{\varepsilon \mathbf{b}} \mathcal{B}_{\varepsilon \mathbf{e}}$ inside $S$ produces a Thomas rotation $\mathbf{R}_{\varepsilon \mathbf{b}}$. This rotation manifests itself in the equation of motion (48) as a velocity rotating 'magnetic' force $-q\mathbf{b} \times \mathbf{u}$, just as the rotation of the turntable manifested itself as a Coriolis force $-2m\boldsymbol{\omega} \times \dot{\mathbf{x}}$ in (12). There is no factor 2 in $-q\mathbf{b} \times \mathbf{u}$ because the Thomas rotation here acts only on velocities, not on positions, so that no tangential velocity comes into play (hence there is no 'centrifugal' force either). [14]

We started with a particle at rest to be able to apply Newton's law, and therefrom derive the relativistic equation of motion. But once the latter is known, we may start with an inertial frame $S$ wherein $\mathbf{b} = 0$, but the particle can have any velocity $\mathbf{u}$. The equation of motion in $S$ is then $U(\tau + \delta\tau) = \mathcal{B}_{\varepsilon \mathbf{e}} U(\tau)$, by (46). If $S$ has velocity $\mathbf{v}$ in $S'$, then similarly to (46):

$$U'(\tau + \delta\tau) = R_{\varepsilon \mathbf{b}'} \mathcal{B}_{\varepsilon \mathbf{e}'} U'(\tau), \qquad \mathbf{e}' = \mathbf{K}_v^{-1} \mathbf{e}, \qquad \mathbf{b}' = -c^{-2}\gamma_v \mathbf{e} \times \mathbf{v} \tag{49}$$

Again, $\mathbf{b}'$ expresses Thomas rotations, arising when one Lorentz transforms boosts induced by an electric field (on a particle of any velocity), in a frame $S$ wherein $\mathbf{b} = 0$. If no such frame exists, as in the vicinity of an electrically neutral wire carrying a net current (since $\mathbf{e}^2 - c^2\mathbf{b}^2$ is invariant), still $\mathbf{b}'$ can be viewed as a *superposition* of elementary fields $\mathbf{b}'_i$, each due to an elementary charge $q_i$, and enacting a Thomas rotation arising from Lorentz transforming to $S'$ boosts induced by $\mathbf{e}_i$ (due to $q_i$) *inside* the rest frame $S_i$ of $q_i$, wherein $\mathbf{b}_i = 0$.

---

[14] Note that a particle at *rest* in $S_0$ would not feel a field $\mathbf{b}_0 \neq 0$, so that one could also write (45) as $U_0(\tau + \delta\tau) = \mathcal{R}_{\varepsilon \mathbf{b}_0} \mathcal{B}_{\varepsilon \mathbf{e}_0} U_0(\tau)$, with *any* $\mathbf{b}_0$. This would modify $\mathbf{e}$ and $\mathbf{b}$ in $S$, but not the total $\mathbf{f}$ in (48). So for a *single* particle of velocity $\mathbf{u}$, it is largely arbitrary how the force $\mathbf{f}$ acting on it is broken up into $\mathbf{f} = \mathbf{e} - \mathbf{b} \times \mathbf{u}$. However, by observing particles of various velocities $\mathbf{u}$, one would find that a unique $\mathbf{e}, \mathbf{b}$ makes (48) valid for all $\mathbf{u}$. This is analogous to the turntable case, see footnote 8.

## 8 The field equations

The preceding derivation of the relativistic equation of motion, and the emergence of the magnetic force, did not depend on the *form* of the Newtonian force field $\mathbf{e}_0(\mathbf{x}_0,t_0)$ in $S_0$. We will now see that if $\mathbf{e}_0(\mathbf{x}_0)$ is a static Coulomb field in $S_0$, then Lorentz transforming to another inertial frame $S$ yields Maxwell's equations, so that Faraday's magnetic induction, the other facet of magnetism besides the Lorentz force, also emerges.

First let us rewrite the electrostatic equations (18) in 4-dimensional notation as

$$\begin{pmatrix} 0 & -\mathbf{e}_0 \\ \mathbf{e}_0^{\mathsf{T}} & 0 \end{pmatrix} \begin{pmatrix} \overleftarrow{\tilde{\nabla}}_0 \\ \overleftarrow{\tilde{\partial}}_{t_0} \end{pmatrix} = \begin{pmatrix} 0 \\ \rho_0 \end{pmatrix}, \qquad \begin{pmatrix} \mathbf{J}_{\mathbf{e}_0} & 0 \\ 0 & 0 \end{pmatrix} \begin{pmatrix} \overleftarrow{\tilde{\nabla}}_0 \\ \overleftarrow{\tilde{\partial}}_{t_0} \end{pmatrix} = 0 \qquad (50)(a,b)$$

where the arrows on the differential operators signify that they act towards the *left*, so that $\mathbf{J}_{\mathbf{a}}\overleftarrow{\tilde{\nabla}} = \mathbf{a} \times \overleftarrow{\tilde{\nabla}} = -\nabla \times \mathbf{a}$. To make the calculations more transparent, it is convenient here to introduce the antisymmetric contravariant tensor $\mathcal{F} = \mathcal{G}g^{-1}$ and its dual, see (44):

$$\mathcal{F}_{\mathbf{e},\mathbf{b}} \equiv \mathcal{G}_{\mathbf{e},\mathbf{b}}g^{-1} = \begin{pmatrix} -\mathbf{J}_{c^2\mathbf{b}} & -\mathbf{e} \\ \mathbf{e}^{\mathsf{T}} & 0 \end{pmatrix}, \qquad \bar{\mathcal{F}}_{\mathbf{e},\mathbf{b}} \equiv \mathcal{F}_{\mathbf{b},-\mathbf{e}/c^2} = \begin{pmatrix} \mathbf{J}_{\mathbf{e}} & -\mathbf{b} \\ \mathbf{b}^{\mathsf{T}} & 0 \end{pmatrix} \qquad (51)$$

Then, in view of (27), the electrostatic eqs (50) can be written as

$$\mathcal{F}_{\mathbf{e}_0,0}\overleftarrow{\tilde{\partial}}_0 = \begin{pmatrix} 0 \\ \rho_0 \end{pmatrix}, \qquad \bar{\mathcal{F}}_{\mathbf{e}_0,0}\overleftarrow{\tilde{\partial}}_0 = 0 \qquad (52)$$

If now $S_0$ has velocity $\mathbf{v}_0$ in $S$, then acting $\mathcal{B}_{\mathbf{v}_0}$ on the left, and inserting $\mathcal{B}_{\mathbf{v}_0}^{\mathsf{T}}\mathcal{B}_{\mathbf{v}_0}^{-1\mathsf{T}} = 1$, yields in view of (27), (41) and (44):

$$\mathcal{F}_{\mathbf{e},\mathbf{b}}\overleftarrow{\tilde{\partial}} = \begin{pmatrix} \mathbf{j} \\ \rho \end{pmatrix}, \qquad \bar{\mathcal{F}}_{\mathbf{e},\mathbf{b}}\overleftarrow{\tilde{\partial}} = 0, \qquad \rho = \gamma_{v_0}\rho_0, \qquad \mathbf{j} = \gamma_{v_0}\mathbf{v}_0\rho_0 = \mathbf{v}_0\rho \qquad (53)(a,b,c,d)$$

$$\mathbf{e} = \mathbf{K}_{v_0}^{-1}\mathbf{e}_0, \qquad \mathbf{b} = c^{-2}\gamma_{v_0}\mathbf{v}_0 \times \mathbf{e}_0 \qquad (54)$$

['duality' ensures that $\mathbf{b}$ is the same in both eqs (53)(a,b); and (c,d) express charge conservation plus Lorentz contraction]. Eqs (53)(a,b) are Maxwell's equations:

$$c^2\nabla \times \mathbf{b} - \partial_t\mathbf{e} = \mathbf{j}, \qquad \nabla \cdot \mathbf{e} = \rho, \qquad \nabla \cdot \mathbf{b} = 0, \qquad \nabla \times \mathbf{e} + \partial_t\mathbf{b} = 0 \qquad (55)(a,b,c,d)$$

Since Eqs (55) do not refer explicitly to $\mathbf{v}_0$, one can set, similarly to (22):

17$$\rho = \sum_i \gamma_i \rho_i, \qquad \mathbf{j} = \sum_i \gamma_i \mathbf{v}_i \rho_i, \qquad \mathbf{e} = \sum_i \mathbf{K}_{\mathbf{v}_i}^{-1} \mathbf{e}_i, \qquad \mathbf{b} = c^{-2} \sum_i \gamma_i \mathbf{v}_i \times \mathbf{e}_i \qquad (56)$$

Thus, Lorentz transforming the electrostatic field equations (18) in $S_0$ yields Maxwell's equations in $S$: the electric field $\mathbf{e}_0$ in $S_0$ produces $\mathbf{b}$ in $S$, as again enacting a Thomas rotation, and this translates *inside $S$* as $\mathbf{e}$ and $\mathbf{b}$ *inducing one another*. Symmetrically, if $S$ has velocity $\mathbf{v}$ in $S'$, and we now let $\mathbf{e} = 0$, then $\mathbf{b}$ in $S$ produces $\mathbf{e}'$ in $S'$ as an 'anti-Coriolis' counter-term, and this translates again inside $S'$ as $\mathbf{e}'$ and $\mathbf{b}'$ [given by (41)] inducing one another. This reciprocal induction enables the two fields to co-propagate in space. Indeed, recall that (55)(a,d) combine to yield $(\partial_t^2 - c^2 \nabla^2)(\mathbf{e}, \mathbf{b}) = 0$ in empty space, implying that electromagnetic disturbances travel at speed $c$ [2-12].

## 9 The Galilean limit

The limit $c \to \infty$ of Maxwell's eqs (55)-(56) yields the Galilean eqs (20)-(22), *provided* one recalls that $\mathbf{b}$ (due to Thomas rotations) is of order $c^{-2}$, so that

$$\lim_{c \to \infty} \mathbf{b} = 0, \qquad \lim_{c \to \infty} c^2 \mathbf{b} = \tilde{\mathbf{b}} \equiv \sum_i \mathbf{v}_i \times \mathbf{e}_i \qquad (57)$$

Also, the limit of the relativistic equation of motion (48)(a) is $d\mathbf{p}/dt = q\mathbf{e}$ with *no* magnetic force, ensuring consistency with the field equations. In view of (31), (39) and (51), we have [since $\gamma_\mathbf{v} \to 1$ and $\mathbf{K}_\mathbf{v} \to \mathbf{1}$ as $c \to \infty$]:

$$B_\mathbf{v} \equiv \lim_{c \to \infty} \mathcal{B}_\mathbf{v} = \begin{pmatrix} \mathbf{1} & \mathbf{v} \\ 0 & 1 \end{pmatrix}, \qquad G_{\mathbf{e},\mathbf{b}} \equiv \lim_{c \to \infty} \mathcal{G}_{\mathbf{e},\mathbf{b}} = \begin{pmatrix} 0 & \mathbf{e} \\ 0 & 0 \end{pmatrix} \qquad (58)(a,b)$$

$$F_{\mathbf{e},\tilde{\mathbf{b}}} \equiv \lim_{c \to \infty} \mathcal{F}_{\mathbf{e},\mathbf{b}} = \begin{pmatrix} -\mathbf{J}_{\tilde{\mathbf{b}}} & -\mathbf{e} \\ \mathbf{e}^\mathsf{T} & 0 \end{pmatrix}, \qquad \bar{F}_{\mathbf{e},\mathbf{b}} \equiv \lim_{c \to \infty} \bar{\mathcal{F}}_{\mathbf{e},\mathbf{b}} = \begin{pmatrix} \mathbf{J}_\mathbf{e} & 0 \\ 0 & 0 \end{pmatrix} \qquad (59)(a,b)$$

Note that Eqs (20) can be obtained directly [bypassing (19)] [15] in the same manner as Eqs (55), by transforming (50) with the Galilean boost $B_\mathbf{v}$. Here however, $G_{\mathbf{e},\mathbf{b}}$ or $G_{\mathbf{e},\tilde{\mathbf{b}}} \equiv \begin{pmatrix} -\mathbf{J}_{\tilde{\mathbf{b}}} & \mathbf{e} \\ 0 & 0 \end{pmatrix}$, and $F_{\mathbf{e},\tilde{\mathbf{b}}}$, and $\bar{F}_{\mathbf{e},\mathbf{b}}$, are unrelated objects, because as $c \to \infty$, the identity $\Lambda^{-1} = g^{-1} \Lambda^\mathsf{T} g$ in (25), and the relation $\mathcal{T} \equiv Tg^{-1}$ between *tensors* $\mathcal{T} \to \Lambda \mathcal{T} \Lambda^\mathsf{T}$ and *transformations* $T \to \Lambda T \Lambda^{-1}$, and the duality (44), break down together with the metric $g$. Indeed, transforming $G_{\mathbf{e},\tilde{\mathbf{b}}}$, $F_{\mathbf{e},\tilde{\mathbf{b}}}$, and $\bar{F}_{\mathbf{e},\mathbf{b}}$, with $B_v$, yields different expressions for $\mathbf{e}', \mathbf{b}'$, namely Eqs (15) ($\mathbf{e}' = \mathbf{e} + \tilde{\mathbf{b}} \times \mathbf{v}$, $\tilde{\mathbf{b}}' = \tilde{\mathbf{b}}$), Eqs (23) ($\mathbf{e}' = \mathbf{e}$, $\tilde{\mathbf{b}}' = \tilde{\mathbf{b}} + \mathbf{v} \times \mathbf{e}$), and ($\mathbf{e}' = \mathbf{e}$, $\mathbf{b}' = \mathbf{b} = 0$), respectively (see Appendix B.3). Observe, nonetheless, that since $\tilde{\mathbf{b}} \times \mathbf{v}$ and $\mathbf{v} \times \mathbf{e}$ both appear in the Galilean expressions – even if in an incoherent way – so *must* they, in a *coherent* way, in the Lorentzian case, since transforming $\mathcal{G}_{e,b}$, $\mathcal{F}_{e,b}$, or $\bar{\mathcal{F}}_{e,b}$, with $\mathcal{B}_v$, necessarily produces the *same* $\mathbf{e}', \mathbf{b}'$. So in a sense, all aspects of magnetism are 'rooted' in Galilean kinematics (see also Footnote 12).

---

[15] Eqs (19) follow from putting $\tilde{\partial}_0 = B_{\mathbf{v}_0}^\mathsf{T} \tilde{\partial} = (\nabla, \partial_t + \mathbf{v}_0 \cdot \nabla)$ in (50).

## 10 Lines of force

> *In my mind's eye, Horatio.*    Hamlet

> *A method which, in Faraday's hands, was far more powerful is that in which he makes use of those lines of magnetic force which were always in his mind's eye, […] and the delineation of which by means of iron fillings he rightly regarded as a most valuable aid to the experimentalist. Faraday looked on these lines as expressing, not only by their direction that of the magnetic force, but by their number and concentration the intensity of that force.*    J. C. Maxwell [24]

The field $\mathbf{e}$ is often represented by Faraday lines of force, of density (number of lines piercing a unit transverse area) proportional to $\mathbf{e}$. If $\mathbf{b} = 0$ in (41), then by (37) and (30)(c):

$$\mathbf{e}' \cdot d\mathbf{s}' = (\mathbf{K}_v^{-1} \mathbf{e}) \cdot (\mathbf{K}_v d\mathbf{s}) = \mathbf{e} \cdot d\mathbf{s} \qquad (60)$$

This tells us that the number of $\mathbf{e}$ lines through *any* surface element $d\mathbf{s}$ is invariant, implying that these lines of force Lorentz contract as if 'etched' in space. If $\mathbf{b} \neq 0$, then a (Galilean) 'anti-Coriolis' counterterm $\gamma_v \mathbf{b} \times \mathbf{v}$ superposes on the contracted lines. Likewise, the $\mathbf{b}$ lines of force Lorentz contract, and a (dual) 'Coriolis' term $c^{-2} \gamma_v \mathbf{v} \times \mathbf{e}$, embodying Thomas rotations, gets added on. So each term in Eqs (3) or (41) has a clear kinematical meaning.

As a famous special case, discovered by Heaviside in 1888 [19, 20], the electric lines of force of a point charge of uniform velocity $\mathbf{u}$ are just the isotropic radial lines of the charge at rest, but compressed by $\gamma_u^{-1}$ along $\mathbf{u}$. This has generally been considered surprising.

> *Suppose you were to draw on a piece of paper the field lines for a charge at rest, and then set the picture to travelling with the speed $v$. Then, of course, the whole picture would be compressed by the Lorentz contraction […] The* miracle *of it is that the picture you would see as the page flies by would still represent the field lines of the point charge.*    Feynman [2f]

On the contrary, we see here that everything is exactly as should be, since Lorentz contraction applies to space *with everything in it*.

## 11 The equation of motion in a different guise

Let us write Eq (47) as follows [noting that $(\mathbf{b} \times \mathbf{u}) \cdot \mathbf{u} = 0$]:

$$\frac{d}{d\tau}U(\tau) = \frac{\gamma_u}{m}\begin{pmatrix} \mathbf{f} \\ c^{-2}\mathbf{f}\cdot\mathbf{u} \end{pmatrix} = \frac{\gamma_u}{m}\begin{pmatrix} 0 & \mathbf{f} \\ c^{-2}\mathbf{f}^{\mathsf{T}} & 0 \end{pmatrix}\begin{pmatrix} \mathbf{u} \\ 1 \end{pmatrix}, \qquad \mathbf{f} = q(\mathbf{e} - \mathbf{b}\times\mathbf{u}) \tag{61}$$

Or equivalently, since $\delta\tau = \delta t/\gamma_u$, and defining an infinitesimal velocity $\delta\mathbf{w} \equiv \mathbf{f}\delta\tau/m$:

$$U(t+\delta t) = \mathcal{B}_{\delta\mathbf{w}}U(t), \qquad \delta\mathbf{w} \equiv \mathbf{f}\delta\tau/m = \mathbf{f}\delta t/m_u, \qquad m_u \equiv \gamma_u m \tag{62}$$

Let us denote here relativistic velocity addition by $\mathbf{v} \oplus \mathbf{u} = \mathcal{B}_\mathbf{v}^\phi \mathbf{u}$ ($\phi$ signifies 'linear fractional map', see Appendix C): while the notation $\mathbf{v} \oplus \mathbf{u}$ conveys the notion of 'addition', $\mathcal{B}_\mathbf{v}^\phi \mathbf{u}$ conveys that of 'transformation'. We may then write the spatial parts of (46) and (62) as:

$$\mathbf{u}(t+\delta t) = \mathcal{B}_{\mathbf{u}(t)}^\phi \delta\mathbf{w}_0 = \mathcal{B}_{\delta\mathbf{w}}^\phi \mathbf{u}(t), \qquad \delta\mathbf{w}_0 \equiv \mathbf{f}_0 \delta t/m_u, \qquad \mathbf{f}_0 = q\mathbf{e}_0 \tag{63}$$

where $\mathbf{u}(t+\delta t) = \mathcal{B}_{\mathbf{u}(t)}^\phi \mathbf{u}_0(t+\delta t) = \mathcal{B}_{\mathbf{u}(t)}^\phi \delta\mathbf{w}_0$ [eq (46)] is just Newton's law in the instantaneous rest frame of the particle. The equation $\mathbf{u}(t+\delta t) = \mathcal{B}_{\delta\mathbf{w}}^\phi \mathbf{u}(t)$ [eq (62)] (which one would use for a step by step numerical integration of the motion) suggests that a particle of velocity $\mathbf{u}$ may be viewed as a 'quasi-particle' of 'rest mass' $m_u = \gamma_u m_0$: an impulse $\mathbf{f}\delta t$ gives to that "particle-of-velocity $\mathbf{u}$" object an infinitesimal velocity $\delta\mathbf{w}$, relative to which the 'bare' particle *maintains* its velocity $\mathbf{u}$ (so that the new velocity of the particle in the laboratory is $\mathcal{B}_{\delta\mathbf{w}}^\phi \mathbf{u} = \delta\mathbf{w} \oplus \mathbf{u}$). Thereby it is possible to treat a bunch of particles of various velocities as an 'object' of 'rest mass' equal to the sum of the individual relativistic masses [6]. This possibility, obvious in the Galilean case, has of course been crucial to the development of physics.

Note that Newton's equation of motion (17) may also be written in the forms (62)-(63), but with Galilean boosts $B_\mathbf{v}$ (instead of Lorentzian boosts $\mathcal{B}_\mathbf{v}$), that is:

$$\mathbf{u}(t+\delta t) = B_{\mathbf{u}(t)}^\phi \delta\mathbf{w} = B_{\delta\mathbf{w}}^\phi \mathbf{u}(t), \qquad \delta\mathbf{w} = \mathbf{f}\delta t/m, \qquad \mathbf{f} = q(\mathbf{e}-\mathbf{b}\times\mathbf{u}) \tag{64}$$

That $B_\mathbf{u}^\phi \delta\mathbf{w} = B_{\delta\mathbf{w}}^\phi \mathbf{u} = \mathbf{u} + \delta\mathbf{w}$ is obvious [recall that here $\mathbf{f}' = \mathbf{f} = \mathbf{f}_0$, see after eq (17)]. By contrast, relativistic 'velocity addition' $\mathbf{u} \oplus \mathbf{v} = \mathcal{B}_\mathbf{u}^\phi \mathbf{v}$ is non-commutative, $\mathbf{u} \oplus \mathbf{v} \neq \mathbf{v} \oplus \mathbf{u}$, for the two sides differ by a Thomas rotation. It follows that $\mathcal{B}_\mathbf{u}^\phi \delta\mathbf{w}_0$ can equal $\mathcal{B}_{\delta\mathbf{w}}^\phi \mathbf{u}$ only if $\mathbf{f}$ and $\mathbf{f}_0$ differ by a transverse term (i.e., a magnetic force) enacting the Thomas rotation.

## 12 Conclusion

That Lorentz covariance implies magnetic forces, alongside electric forces, has been argued in various ways. One way (recalled in section 1) is to consider the radial force exerted by a current-carrying wire on a test charge moving parallel to it [2, 8, 9]. This argument, however, does not work if the test charge moves transversally to the wire. Also, it gives no clue as to why the magnetic force has its strange, Coriolis-like, velocity-rotating character.

Now, due to the spatial anisotropy (expressed by $\mathbf{K_v}$) forced into Lorentz boosts by space-time interval invariance, products of non-collinear Lorentz boosts produce Thomas rotations (intimately tied to the non-commutativity of velocity 'addition'). It is noteworthy that Silberstein [10] ended his discussion of Thomas rotations with the remark:

> *"We have touched the six-parameter Lorentz group only to elucidate the question of successive transformations, as intimately connected with the composition of velocities. But henceforth we shall hardly need it any more. In fact, our previous transformation $\Lambda_v$ without any rotation of the space-framework, will be found sufficient for all physical purposes."*

Indeed, most textbooks on relativity hardly ever mention Thomas-Wigner rotations. Yet they are the unsung culprit behind magnetism.

By writing Newton's equation of motion for a charged particle *at rest* in a form displaying explicitly that an electric field acts on its velocity as a boost generator, and Lorentz transforming, we found that magnetic forces enact Thomas rotations. Moreover, Lorentz transforming the Coulomb electrostatic field equations yields Maxwell's equations, so that magnetic induction, the other facet of magnetism, also emerges, again as an effect of Thomas rotations.

But it is true that in general there is no need to mention Thomas rotations. Still, from a conceptual point of view, it is nice to understand wherefrom comes the peculiar Coriolis-like nature of magnetic forces (and, also, that electromagnetic fields transform as second-rank tensors simply because they parametrize generators of Lorentz transformations).

There are of course important differences between magnetic and Coriolis forces. In particular, Coriolis forces are proportional to the masses $m$ of particles, while magnetic forces are independent of $m$, being caused by Thomas rotations proportional to infinitesimal boosts $\varepsilon\mathbf{e} = q\mathbf{e}\delta\tau/m$. Thus, particles of equal velocities but different masses disperse in a magnetic field, but respond in unison to Coriolis forces (as they do to a gravitational field).

Thus, Lorentz boosts are not so innocent: They create rotations, hence 'fictive' forces, ascribed to magnetic fields. Quantum mechanics has the habit of preying on 'fictive' classical stuff and making it 'real': For instance, magnetic potentials appear quite 'real' in the Bohm-Aharonov effect [2d]; another example (perhaps) is the magnetic field associated with the spin of a 'point' particle with (apparently) no internal structure, like an electron. [16]

Realizing that magnetism is a kind of 'Coriolis' effect changes nothing, except maybe to make more palatable the fact that magnetic monopoles have not shown up (yet) – but in no way forbids them [16-18, 9]. Heaviside, for one, liked to write Maxwell's equations in a symmetric 'duplex form', including magnetic charges and currents, thereby *"throwing all potentials overboard"* [15a]: Classically, indeed, magnetic potentials are incompatible with magnetic charges, as Dirac reminds us in his groundbreaking article on magnetic monopoles [16].

> *Elementary classical theory allows us to formulate equations of motion for an electron in the field produced by an arbitrary distribution of electric charges and magnetic poles. If we wish to put the equations of motion in the Hamiltonian form, however, we have to introduce the electromagnetic potentials, and this is possible only when there are no isolated magnetic poles.*

And since *quantum* mechanics *requires* these potentials for its formulation, it seemed that magnetic monopoles were excluded once and for all.

> *Quantum mechanics, as it is usually established, is derived from the Hamiltonian form of the classical theory and therefore is applicable only when there are no isolated magnetic poles.*

Yet Dirac went on to show that they nevertheless can (and, decently, should) exist.

> *The object of the present paper is to show that quantum mechanics does not really preclude the existence of isolated magnetic poles. On the contrary, the present formalism of quantum mechanics […] leads inevitably to wave equations whose only physical interpretation is the motion of an electron in the field of a single pole […] one would be surprised if Nature had made no use of it.*     Dirac 1931 [16]

Moreover, the mere existence of a single magnetic monopole would explain the quantisation of electric charge [16-18].

> *The mere existence of one pole of strength $g$ would require all electric charges to be quantized in units of $\hbar c / 2g$ […]. The quantization of electricity is one of the most fundamental and striking features of atomic physics, and there seems to be no explanation for it apart from the theory of poles. This provides some grounds for believing in the existence of these poles.*     Dirac 1948 [17]

But *"none has been found"* (Feynman – see Footnote 1).

---

[16] Feynman, in his *Lectures*, seemed to visualize electrons as tiny charged gyroscopes [2e].

Electric and magnetic fields **e** and **b** act as Lorentz generators on the velocities of charged particles. So by considering how an infinitesimal Lorentz transformation *inside* a given inertial frame transforms to another inertial frame, one can understands *kinematically* the way **e** and **b** Lorentz transform: as should be, electric and magnetic lines of force *Lorentz contract* as if 'etched' in space, while '(anti)Coriolis' terms due to rotations get added on.

The contraction of the radial electric lines of force of a moving point charge provides a nice way of understanding electromagnetic radiation (due to J. J. Thomson [25]): If the charge suddenly accelerates to a new velocity, then transverse kinks are created in these lines of force, since the contraction is different before and after the acceleration. As these kinks propagate outwards, their separations grow like the radial distance $r$, hence their density decreases like $r^{-1}$. In contrast, the density of *radial* lines (hence the radial electric field) decreases like $r^{-2}$. The kinks $\sim r^{-1}$ constitute the radiation field:

> *We have a little piece of field which is travelling through space all by itself. The fields have "taken off"; they are propagating freely through space, no longer connected in any way with the source. The caterpillar has turned into a butterfly!*     Feynman [2f]

Thus, a maximum speed $c$ for signals implies Lorentz covariance; implying in turn that electric fields produce (due to Thomas rotations) 'Coriolis' magnetic fields, such that these two fields, when changing in time, induce one another, enabling them to co-propagate at speed $c$.

> *How can this bundle of electric and magnetic fields maintain itself? The answer is: by the combined effects of the Faraday law $\nabla \times \mathbf{e} = -\partial_t \mathbf{b}$, and the new term of Maxwell, $c^2 \nabla \times \mathbf{b} = \partial_t \mathbf{e}$. They cannot help maintaining themselves. Suppose the magnetic field were to disappear. There would be a changing magnetic field which would produce an electric field. If this electric field tries to go away, the changing electric field would create a magnetic field back again* […] *They must go on forever. They maintain themselves in a kind of a dance – one making the other.*     Feynman [2f]

Recall from Eq (20)(a) that Maxwell's 'new term' $\partial_t \mathbf{e}$ (in Ampère's law $c^2 \nabla \times \mathbf{b} - \partial_t \mathbf{e} = \mathbf{j}$) has a humble Galilean origin: [17] For if, within Galilean relativity (wherein Thomas rotations and hence magnetism do not occur), a velocity-rotating magnetic field **b** is put in 'by hand', then a Galilean (hence also Lorentzian) boost by velocity **v** creates an 'anti-Coriolis' electric field, in order to undo the rotation of **v** by **b**; and this 'anti-Coriolis' field entails the term $\partial_t \mathbf{e}$.

---

[17] So does, in a sense, Faraday's law, by 'duality', or as described in Footnote 12.

The above 'dance' between 'real' and 'Coriolis' fields, in electromagnetic radiation, has more earthly precedents: Imagine a frictionless spinning top: as it tips under the pull of gravity, it rotates about a horizontal axis. This rotation induces Coriolis forces on the rapidly moving atoms of the spinning top, causing it to precess [23]. This new rotation, about a vertical axis, in turn induces Coriolis forces which push the top back towards the vertical. Whence the nutation of the top. In this endless up-down oscillation, gravitational and Coriolis forces alternatively reanimate one another *"in a kind of a dance"*. Riding a bicycle (*"the salvation of the body"* - Fitzgerald [15g]) is also a 'dance' balancing gravitational and 'fictive' inertial forces [26].[18]

We noted that the relativistic equation of motion allows to treat a bunch of particles of varied velocities as an 'object' of rest mass equal to the sum of the individual relativistic masses. Indeed, physics would be in a sorry state had it not been possible to treat agregates of elementary particles (whatever that means) as 'elementary' objects. As Bohm stressed [6]:

> *The property possessed by bulk matter – being capable alternatively of analysis into parts or treatment as a single whole – is a general feature of the world. This feature must therefore be implied by any proposed set of laws of mechanics.*     David Bohm [6]

This demand in fact suffices [6] to deduce the relativistic equation of motion (48). As often in physics, 'simplicity' is here an unreasonably effective guide:

> *One has no other guide for guessing the right laws than the ideal of simplicity.*    Max Born [5]

This is not always the case in science:

> *Elegance and simplicity are, in biology, dangerous guides. Biologists must constantly remind themselves that what they see was not designed but evolved.*     F. H. Crick [27]

**Acknowledgements**: I thank Gonzalo Reyes for interesting me in the nature of forces in special relativity, and for numerous discussions.---

[18] Leaning the bicycle to the right or left under you causes Coriolis forces on the spinning wheels, which turn them right or left (see Footnote 9); thus you can drive freehand. In case a sudden right turn is called for, motorcyclists are trained to 'countersteer', i.e., give a jolt turning the front wheel to the *left*: centrifugal forces, and Coriolis forces on the wheels, then make the motorcycle lean to the *right*, thereby inducing other Coriolis forces which (violently) turn the wheels to the right, whence a sudden right turn [26].

**Appendix A   Connection with tensor notation**

Block matrix notation allows one to keep track simultaneously of Lorentz covariance and separate space and time aspects. We here relate (with two-way arrows $\leftrightarrow$) matrix equations with the corresponding tensor equations.

Vectors $X \leftrightarrow X^\mu$ transform as $X' = \Lambda X \leftrightarrow X'^\mu = \Lambda^\mu{}_\nu X^\nu$, where $\Lambda \leftrightarrow \Lambda^\mu{}_\nu$ satisfies (25)(a): $\Lambda^\mathsf{T} g \Lambda = g \leftrightarrow \Lambda^\alpha{}_\mu g_{\alpha\beta} \Lambda^\beta{}_\nu = g_{\mu\nu}$, where $g \leftrightarrow g_{\mu\nu}, g^{-1} \leftrightarrow g^{\mu\nu}$ is the metric. If $\tilde{Y} = gY \leftrightarrow Y_\mu = g_{\mu\nu} Y^\nu$, then $\tilde{Y}^\mathsf{T} X \leftrightarrow Y_\mu X^\mu$ is invariant. Also, $\tilde{\partial} \leftrightarrow \partial_\mu$, $\partial \leftrightarrow \partial^\mu$. Treating the Lorentz transformation matrix $\Lambda \leftrightarrow \Lambda^\mu{}_\nu$ as a 'tensor', we can write $\Lambda^{-1\mathsf{T}} = g \Lambda g^{-1} \leftrightarrow \Lambda_\mu{}^\nu = g_{\mu\alpha} \Lambda^\alpha{}_\beta g^{\beta\nu}$ [28], so that $\tilde{Y}' = \Lambda^{-1\mathsf{T}} Y \leftrightarrow Y'_\mu = \Lambda_\mu{}^\nu Y_\nu$. The electromagnetic tensor has the Lorentz-generator form $\mathcal{G} \leftrightarrow \mathcal{G}^\mu{}_\nu$, and the antisymmetric form $\mathcal{F} \equiv \mathcal{G} g^{-1} \leftrightarrow \mathcal{F}^{\mu\nu} = \mathcal{G}^\mu{}_\alpha g^{\alpha\nu}$. These transform as

$$\mathcal{G}' = \Lambda \mathcal{G} \Lambda^{-1} \leftrightarrow \mathcal{G}'^\mu{}_\nu = \Lambda^\mu{}_\alpha \mathcal{G}^\alpha{}_\beta \Lambda_\nu{}^\beta, \qquad \mathcal{F}' = \Lambda \mathcal{F} \Lambda^\mathsf{T} \leftrightarrow \mathcal{F}'^{\mu\nu} = \Lambda^\mu{}_\alpha \mathcal{F}^{\alpha\beta} \Lambda^\nu{}_\beta \qquad (65)$$

In explicit $4 \times 4$ matrix notation:

$$\mathcal{G}_{\mathbf{e,b}} = \begin{pmatrix} 0 & b_z & -b_y & e_x \\ -b_z & 0 & b_x & e_y \\ b_y & -b_x & 0 & e_z \\ c^{-2} e_x & c^{-2} e_y & c^{-2} e_z & 0 \end{pmatrix}, \qquad \mathcal{F}_{\mathbf{e,b}} = \begin{pmatrix} 0 & c^2 b_z & -c^2 b_y & -e_x \\ -c^2 b_z & 0 & c^2 b_x & -e_y \\ c^2 b_y & -c^2 b_x & 0 & -e_z \\ e_x & e_y & e_z & 0 \end{pmatrix} \qquad (66)$$

Eqs (53) are $\partial^\nu \mathcal{G}^\mu{}_\nu = j^\mu$ and $\partial^\nu \bar{\mathcal{G}}^\mu{}_\nu = 0$, where $\bar{\mathcal{G}}$ is the dual electromagnetic tensor, obtained by doing $\mathbf{e} \to \mathbf{b}$, $c^2 \mathbf{b} \to -\mathbf{e}$, or $\bar{\mathcal{F}}_{\mu\nu} = \varepsilon_{\mu\nu\alpha\beta} \mathcal{F}^{\alpha\beta}$, where $\varepsilon_{\mu\nu\alpha\beta}$ is the totally antisymmetric tensor. This last tensor relation makes obvious the duality (44).

**Appendix B   Explicit computations**

B.1 *Derivation of Eqs* (41): It is a bit simpler to work with $\mathcal{F} = \mathcal{G} g^{-1}$ than with $\mathcal{G}$. Using $\mathbf{K}_{-\mathbf{v}} = \mathbf{K}_\mathbf{v}^\mathsf{T} = \mathbf{K}_\mathbf{v}$, and (10), we first compute:

$$\mathcal{B}_\mathbf{v} \mathcal{F}_{\mathbf{e},0} \mathcal{B}_\mathbf{v}^\mathsf{T} = \gamma_v^2 \begin{pmatrix} \mathbf{v}\mathbf{e}^\mathsf{T} \mathbf{K}_\mathbf{v} - \mathbf{K}_\mathbf{v} \mathbf{e}\mathbf{v}^\mathsf{T} & c^{-2} \mathbf{v}\mathbf{e}^\mathsf{T} \mathbf{v} - \mathbf{K}_\mathbf{v} \mathbf{e} \\ \mathbf{e}^\mathsf{T} \mathbf{K}_\mathbf{v} - c^{-2} \mathbf{v}^\mathsf{T} \mathbf{e}\mathbf{v}^\mathsf{T} & c^{-2} (\mathbf{e}^\mathsf{T} \mathbf{v} - \mathbf{v}^\mathsf{T} \mathbf{e}) \end{pmatrix} = \begin{pmatrix} \mathbf{J}_{\gamma_v \mathbf{e} \times \mathbf{v}} & -\mathbf{K}_\mathbf{v}^{-1} \mathbf{e} \\ (\mathbf{K}_\mathbf{v}^{-1} \mathbf{e})^\mathsf{T} & 0 \end{pmatrix} \qquad (67)$$

where we used $\mathbf{v}\mathbf{e}^\mathsf{T} \mathbf{K}_\mathbf{v} - \mathbf{K}_\mathbf{v} \mathbf{e}\mathbf{v}^\mathsf{T} = \gamma_v^{-1} (\mathbf{v}\mathbf{e}^\mathsf{T} - \mathbf{e}\mathbf{v}^\mathsf{T})$, since $\mathbf{K}_\mathbf{v} = \gamma_v^{-1} [\mathbf{1} + (\gamma_v - 1) \boldsymbol{\pi}_\mathbf{v}]$ and $\mathbf{v}\mathbf{e}^\mathsf{T} \boldsymbol{\pi}_\mathbf{v} - \boldsymbol{\pi}_\mathbf{v} \mathbf{e}\mathbf{v}^\mathsf{T} = 0$. Also, we noted that $\mathbf{K}_\mathbf{v} \mathbf{e} - c^{-2} \mathbf{v}\mathbf{e}^\mathsf{T} \mathbf{v} = [\mathbf{K}_\mathbf{v} - (v/c)^2 \boldsymbol{\pi}_\mathbf{v}] \mathbf{e} = \gamma_v^2 \mathbf{K}_\mathbf{v}^{-1} \mathbf{e}$. One deduces $\mathcal{B}_v \mathcal{F}_{0,b} \mathcal{B}_v^T$ by using the duality (44). We then obtain (41) by summing.

**B.2** *Invariants and duality*: Using (66) (with $\mathbf{b}$ along the $x$-axis say), and (10)(b), we get:

$$\mathcal{G}_{e,b}\mathcal{G}_{E,B} = \begin{pmatrix} \mathbf{J_b J_B} + \mathbf{eE}^T/c^2 & -\mathbf{J_b E} \\ -\mathbf{e}^T \mathbf{J_B}/c^2 & \mathbf{e}\cdot\mathbf{E}/c^2 \end{pmatrix} = \begin{pmatrix} \mathbf{Bb}^T - (\mathbf{b}\cdot\mathbf{B})\mathbf{1} + \mathbf{eE}^T/c^2 & -\mathbf{b}\times\mathbf{E} \\ (\mathbf{B}\times\mathbf{e})^T/c^2 & \mathbf{e}\cdot\mathbf{E}/c^2 \end{pmatrix} \quad (68)$$

$$\det \mathcal{G}_{e,b} = -(\mathbf{e}\cdot\mathbf{b}/c)^2, \qquad \mathrm{tr}\,\mathcal{G}_{e,b}^2 = 2(c^{-2}\mathbf{e}^2 - \mathbf{b}^2)$$

(since $\mathrm{tr}\,\mathbf{ab}^T = \mathbf{a}\cdot\mathbf{b}$ and $\mathrm{tr}\,\mathbf{1} = 3$), whence the invariants (43). We also find that

$$\mathcal{G}_{e,b}\bar{\mathcal{G}}_{e,b} = (\mathbf{e}\cdot\mathbf{b}/c^2)\mathbf{1}, \qquad \bar{\mathcal{G}}_{e,b} \equiv \mathcal{G}_{b,-e/c^2} \quad (69)$$

Hence (since $\mathbf{e}\cdot\mathbf{b} = \mathbf{e}'\cdot\mathbf{b}'$): $\mathcal{G}\bar{\mathcal{G}} = \mathcal{G}'\bar{\mathcal{G}}'$ if $\mathcal{G}' \equiv \Lambda\mathcal{G}_{e,b}\Lambda^{-1} = \mathcal{G}_{e',b'}$. Also $\mathcal{G}\bar{\mathcal{G}} = \Lambda\mathcal{G}\bar{\mathcal{G}}\Lambda^{-1} = \mathcal{G}'\Lambda\bar{\mathcal{G}}\Lambda^{-1}$, so that $\mathcal{G}'\bar{\mathcal{G}}' = \mathcal{G}'\Lambda\bar{\mathcal{G}}\Lambda^{-1}$. Thus $\bar{\mathcal{G}}' = \Lambda\bar{\mathcal{G}}\Lambda^{-1}$ if $\det\mathcal{G}' = -(\mathbf{e}\cdot\mathbf{b}/c)^2 \neq 0$, and *also* if $\mathbf{e}\cdot\mathbf{b} = 0$, by continuity. Whence eqs (44).

**B.3** *Galilean transformations in 4-dimensional notation*: We have:

$$\mathbf{G}_{e',\tilde{b}'} = \mathbf{B}_v \mathbf{G}_{e,\tilde{b}} \mathbf{B}_v^{-1} = \begin{pmatrix} \mathbf{1} & \mathbf{v} \\ 0 & 1 \end{pmatrix}\begin{pmatrix} -\mathbf{J}_{\tilde{b}} & \mathbf{e} \\ 0 & 0 \end{pmatrix}\begin{pmatrix} \mathbf{1} & -\mathbf{v} \\ 0 & 1 \end{pmatrix} = \begin{pmatrix} -\mathbf{J}_{\tilde{b}} & \mathbf{e} + \mathbf{J}_{\tilde{b}}\mathbf{v} \\ 0 & 0 \end{pmatrix} \quad (70)$$

whence Eqs (15): $\tilde{\mathbf{b}}' = \tilde{\mathbf{b}}$, $\mathbf{e}' = \mathbf{e} + \tilde{\mathbf{b}}\times\mathbf{v}$. Next:

$$F_{e',\tilde{b}'} = \mathbf{B}_v F_{e,\tilde{b}} \mathbf{B}_v^T = \begin{pmatrix} \mathbf{1} & \mathbf{v} \\ 0 & 1 \end{pmatrix}\begin{pmatrix} -\mathbf{J}_{\tilde{b}} & -\mathbf{e} \\ \mathbf{e}^T & 0 \end{pmatrix}\begin{pmatrix} \mathbf{1} & 0 \\ \mathbf{v}^T & 1 \end{pmatrix} = \begin{pmatrix} -\mathbf{J}_{\tilde{b}} - \mathbf{ev}^T + \mathbf{ve}^T & -\mathbf{e} \\ \mathbf{e}^T & 0 \end{pmatrix} \quad (71)$$

whence Eqs (23): $\tilde{\mathbf{b}}' = \tilde{\mathbf{b}} + \mathbf{v}\times\mathbf{e}$, $\mathbf{e}' = \mathbf{e}$, since $\mathbf{ev}^T + \mathbf{ve}^T = \mathbf{J}_{\mathbf{v}\times\mathbf{e}}$ by (10)(a). Finally:

$$\bar{F}_{e',b'} = \mathbf{B}_v \bar{F}_{e,b} \mathbf{B}_v^T = \begin{pmatrix} \mathbf{1} & \mathbf{v} \\ 0 & 1 \end{pmatrix}\begin{pmatrix} \mathbf{J}_e & 0 \\ 0 & 0 \end{pmatrix}\begin{pmatrix} \mathbf{1} & 0 \\ \mathbf{v}^T & 1 \end{pmatrix} = \begin{pmatrix} \mathbf{J}_e & 0 \\ 0 & 0 \end{pmatrix} \quad (72)$$

whence $\mathbf{e}' = \mathbf{e}$, $\mathbf{b}' = \mathbf{b} = 0$.

**Appendix C   Velocity 'addition' as a linear fractional map**

Let matrices $\mathbf{x}$ and $\mathbf{y}$ have the same number of columns (but not necessarily of rows), and matrices $\mathbf{T}, \mathbf{M}, \mathbf{a}, \mathbf{b}, \mathbf{c}, \mathbf{d}$ be of the proper dimensions for the ensuing equations to make sense (in particular $\mathbf{M}$ is a square matrix). Suppose that $\mathbf{x}$ and $\mathbf{y}$ are linearly related by $\mathbf{x} = \mathbf{Ty}$. Let now $\mathbf{x}$ and $\mathbf{y}$ be linearly mixed into new quantities $\mathbf{x}'$ and $\mathbf{y}'$. We find that

$$\{ \mathbf{x} = \mathbf{T}\mathbf{y}, \quad \mathbf{X'} = \begin{pmatrix} \mathbf{x'} \\ \mathbf{y'} \end{pmatrix} = \mathbf{M} \begin{pmatrix} \mathbf{x} \\ \mathbf{y} \end{pmatrix} = \mathbf{MX} \} \quad \Leftrightarrow \quad \{ \mathbf{x'} = \mathbf{T'}\mathbf{y'}, \quad \mathbf{T'} = \mathbf{M}^\phi \mathbf{T} \} \tag{73}$$

where the *linear fractional map* $\mathbf{M}^\phi$ on matrices is defined by:

$$\mathbf{M} = \begin{pmatrix} \mathbf{a} & \mathbf{b} \\ \mathbf{c} & \mathbf{d} \end{pmatrix} \quad \Leftrightarrow \quad \mathbf{M}^\phi \mathbf{T} = (\mathbf{aT} + \mathbf{b}) \frac{1}{\mathbf{cT} + \mathbf{d}} \tag{74}$$

If now $\mathbf{X''} = \mathbf{M'}\mathbf{X'} = \mathbf{M'}\mathbf{M}\mathbf{X}$, then one has the composition rules

$$\mathbf{M'}^\phi \mathbf{M}^\phi = (\mathbf{M'M})^\phi, \qquad (\mathbf{M}^{-1})^\phi = (\mathbf{M}^\phi)^{-1}, \qquad \mathbf{1}^\phi = \mathbf{1} \tag{75}$$

If $\mathbf{y}$ is an invertible square matrix (e.g., a non-zero scalar), then since $\mathbf{T} = \mathbf{x}\mathbf{y}^{-1}$:

$$\begin{pmatrix} \mathbf{x'} \\ \mathbf{y'} \end{pmatrix} = \mathbf{M} \begin{pmatrix} \mathbf{x} \\ \mathbf{y} \end{pmatrix} \quad \Rightarrow \quad (\mathbf{x'}\mathbf{y'}^{-1}) = \mathbf{M}^\phi (\mathbf{x}\mathbf{y}^{-1}) \tag{76}$$

Applying (73) to $d\mathbf{x} = \mathbf{u}dt$ and $dX' = \mathcal{B}_v dX$, where $dX = (d\mathbf{x}, dt)$, so that $\mathbf{u}$ plays the role of $\mathbf{T}$ in $\mathbf{x} = \mathbf{Ty}$, and $\mathcal{B}_v$ that of $\mathbf{M}$, we get:

$$\mathbf{u'} = \mathbf{v} \oplus \mathbf{u} = \mathcal{B}_v^\phi \mathbf{u} = \frac{\mathbf{K}_v \mathbf{u} + \mathbf{v}}{1 + \mathbf{v} \cdot \mathbf{u}} = \frac{\mathbf{v} + \mathbf{u} - \lambda_v \bar{\pi}_v \mathbf{u}}{1 + \mathbf{v} \cdot \mathbf{u}}, \qquad \lambda_v \equiv 1 - \gamma_v^{-1} \tag{77}$$

Alternatively, one may apply directly (76) (with $\mathbf{x} \to d\mathbf{x}$, $\mathbf{y} \to dt$) to $\mathbf{u} = d\mathbf{x}/dt$.

[1] From a letter to the Michelson Commemorative Meeting of the Cleveland Physics Society, as quoted in R. S. Shankland, Am. J. Phys. **32**, 16 (1964), p. 35. Also quoted in Ref. [8].

[2] R. P. Feynman, R. B. Leighton, M. Sands, *The Feynman lectures on Physics*, Addison-Wesley 1964, Vol II; (a) sect. 1-4; (b) sect. 13-6; (c) sect. 1-4; (d) sect. 15-5; (e) sect. 13-5; (f) sect. 18-4; (g) sect. 26-2.

[3] A. Einstein, Annalen der Physik **17**, 1905. English translation: "On the electrodynamics of moving bodies", in *The principle of relativity*, Dover; also in: *Einstein's miraculous year*, edited by J. Stachel, Princeton U.P., 1998.

[4] W. Pauli, *Theory of Relativity*, Dover, New York 1981.

[5] Max Born, *Einstein's Theory of Relativity*, Dover 1965, page 372.

[6] D. Bohm, *The Special Theory of Relativity*, Benjamin/Cummings, London 1965.

[7] W. Rindler, *Introduction to Special Relativity*, 2nd edition, Clarendon Press, Oxford 1991.

[8] A. P. French, *Special Relativity,* W.W. Norton, New York 1968; (a) p 256; (b) p 243.

[9] D. J. Griffiths, *Introduction to Electrodynamics*, 3rd edition, Prentice Hall, 1999; (a) p 522; (b) p 439; (c) cover pages.

[10] L. Silberstein, *The Theory of Relativity,* Macmillan, London 1914. Second edition 1924. The citations (in sections 1 and 12 of the present paper) are from pages 149 and 169 of the 1924 edition.

[11] M. Kauderer, *Symplectic matrices, first order systems and special relativity,* World Scientific, Singapore 1994, page 194 and after.

[12] W. E. Baylis, *Electrodynamics, a modern geometric approach*, Birkhauser, Boston 1999.

[13] L.H. Thomas, "The motion of the spinning electron", Nature (London) **117**, 514 (1926); "The kinematics of an electron with an axis", Philos. Mag. **3**, 1-22 (1927).

[14] E. P. Wigner, "On unitary representations of the inhomogeneous Lorentz group", Annals of Mathematics **40**, 149-204 (1939). Note that Wigner refered to Silberstein's books.

[15] P. J. Nahan, *Oliver Heaviside*, John Hopkins U. P., 2002; (a) p 128; (b) p109; (c) p 92; (d) p 136, note 85; (e) p 200; (f) p 187; (g) p 268.

[16] P. A. M. Dirac, *Quantised singularities in the electromagnetic field,* Proc. R. Soc. Lond. A **133**, 60-72 (1931) (the citations are from page 71).

[17] P. A. M. Dirac, *The Theory of Magnetic Poles,* Phys. Rev. **74**, 817-830 (1948) (page 817).

[18] See, e.g., A. Zee, *Quantum Field Theory in a Nutshell*, Princeton U.P., 2003, Ch. IV sect. 4.

[19] O. Darrigol, *Electrodynamics from Ampère to Einstein*, Oxford U.P., 2000, page 201.

[20] O. Jefimenko, Am. J. Phys. **62**, 79-85 (1994).

[21] Wu-Ki Tung, *Group Theory in Physics*, World Scientific, Singapore 1985, page 179.

[22] See, e.g., H. Goldstein, C. Poole and J. Safko, *Classical Mechanics,* 3rd edition, Addison-Wesley 2002; T. L. Chow, *Classical Mechanics*, Wiley 1995; V. D. Barger and M. G. Olsson, *Classical Mechanics*, McGraw-Hill 1995.

[23] R. P. Feynman, R. B. Leighton, M. Sands, *The Feynman lectures on Physics*, Addison-Wesley 1964, Vol I, section 20-3.

[24] J. C. Maxwell, *Treatise on electricity and magnetism*, 3rd edition, Dover 1954, vol. 2, article 541.

[25] C. W. Misner, K. S. Thorne, J.A. Wheeler, *Gravitation*, Freeman 1970, sections 4.4 and 41.

[26] J. Fajans, "Steering in bicycles and motorcycles", American Journal of Physics **68**, 654-659 (2000)

[27] Francis Crick, *What mad pursuit (a personal view of scientific discovery)*, Basic books, 1988, p 141.

[28] See, *e.g.*, S. Weinberg, *The Quantum Theory of Fields*, Vol. I, Cambridge U. P. 1995, eq. (2.3.10).